\documentclass[12pt]{report}
\usepackage{amsmath,amssymb}
\usepackage{graphicx}
\usepackage{latexsym}
\usepackage{dcolumn}
\usepackage{enumitem} 
\usepackage{bm}
\title{COSMOLOGY IN TERMS OF THE DECELERATION PARAMETER. PART I}
\author{Yu.L. Bolotin, V.A. Cherkaskiy, O.A. Lemets \\ D.A. Yerokhin and L.G. Zazunov}
\date{\flushright{\small''All of observational cosmology is the search\\ for two numbers: $H_0$ and $q_0$.''\\ Allan Sandage, 1970}}
\begin{document}
\maketitle
\begin{abstract}
In the early seventies, Alan Sandage defined cosmology as the search for two numbers: Hubble parameter ${{H}_{0}}$ and deceleration parameter ${{q}_{0}}$. The first of the two basic cosmological parameters (the Hubble parameter) describes the linear part of the time dependence of the scale factor.  Treating the Universe as a dynamical system it is natural to assume that it is non-linear: indeed, linearity is nothing more than approximation, while non-linearity represents the generic case. It is evident that future models of the Universe must take into account different aspects of its evolution. As soon as the scale factor is the only dynamical variable, the quantities which determine its time dependence must be essentially present in all aspects of the Universe' evolution. Basic characteristics of the cosmological evolution, both static and dynamical, can be expressed in terms of the parameters ${{H}_{0}}$ and ${{q}_{0}}$. The very parameters (and higher time derivatives of the scale factor) enable us to construct model-independent kinematics of the cosmological expansion.

Time dependence of the scale factor reflects main events in history of the Universe. Moreover it is the deceleration parameter who dictates the expansion rate of the Hubble sphere and determines the dynamics of the observable galaxy number variation: depending on the sign of the deceleration parameter this number either grows (in the case of decelerated expansion), or we are going to stay absolutely alone in the cosmos (if the expansion is accelerated).

The intended purpose of the report is reflected in its title --- "Cosmology in terms of the deceleration parameter". We would like to show that practically any aspect of the cosmological evolution is tightly bound to the deceleration parameter.
\end{abstract}
\tableofcontents
\newpage
In the forthcoming parts we plan to cover the following topics:\\
{\bf 5. Deceleration parameter in different cosmological models}
\begin{itemize}
\item SCM
\item Cosmology with power and hybrid expansion laws
\item Cosmological models with constant deceleration parameter (Berman model)
\item Linearly varying deceleration parameter
\end{itemize}
{\bf 6. Kinematic aspects of dynamical forms of the dark energy}
\begin{itemize}
\item Kinematics of scalar fields -- general description
\item Quintessence
\item K-essence
\item Chaplygin gas
\item Phantom
\item Crossing the phantom divide
\item Polytropic model
\item Quintom cosmology
\item Tachyon
\item Chameleon fields
\end{itemize}
{\bf 7. Beyond the standard model}
\begin{itemize}
\item Modified gravity
\item f(T)
\item MOND
\item Models with interaction of the dark components
\item Models with creation of matter
\item Polytropic processes in the expanding Universe
\item Models with cosmological viscosity
\item Fractal cosmology
\item Entropic forces
\item Deceleration parameter in holographic models
\end{itemize}
{\bf 8. History of Universe in terms of the deceleration parameter}
\begin{itemize}
\item Is the transition redshift a new cosmological number?
\item Inflation
\item Inflation, Dark Energy and the Higgs
\item Decelerating Past
\item Fate of the Universe
\end{itemize}
{\bf 9. Dynamical features in terms of the deceleration parameter}
\begin{itemize}
\item Deceleration parameter in anisotropic Universe
\item Deceleration parameter in inhomogeneous Universe
\item Transitive acceleration
\item In vicinity of the singularities
\end{itemize}
{\bf 10. Thermodynamics of an expanding Universe}
\begin{itemize}
\item Thermodynamic description of the expansion kinematics
\item Thermodynamic constraints on the deceleration parameter
\end{itemize}
{\bf 11. Role of quantum effects in cosmological evolution}
\begin{itemize}
\item Dynamics of wave packets
\item Entanglement
\item de Brogle-Bhom corrections
\item Entanglement
\item Deceleration parameter in the loop cosmology
\end{itemize}
{\bf 12. Observational aspects}
\begin{itemize}
\item Reconstruction of Deceleration Parameters from Cosmic Observations
\item A cosmographic test of LCDM
\item Cosmological scalars and the Friedmann equation
\item When did cosmic acceleration start? How fast was the transition?
\item State-finder diagnostic
\item Model-independent methods for exploring the nature of dark energy
\item Parametrizations of the deceleration parameter
\item Evidences in favor of the accelerated expansion of Universe
\item Deceleration parameter and Nobel prize 2011
\item Observations in the angular diameter
\item Sandage-Loeb test
\item A direct measurement of the cosmic acceleration
\item Syunyaev-Zeldovich effect
\item GRB as a standard candle
\item 21 cm
\item Cosmography and CMB
\item Gravity lensing
\item Cosmography beyond Standard Candles and Rulers
\item After PLANCK
\end{itemize}
{\bf 13. Conclusion}

\chapter{Introduction}
In the early seventies, Alan Sandage \cite{sandage} defined Cosmology as the search for two numbers: $H_0$ and $q_0$. It seemed too simple and clear: the main term in form of the Hubble parameter (HP) determined the expansion rate of the Universe and a small correction due to gravity of the matter content was responsible for slow down of the expansion. However the situation drastically changed at the end of the last century.

By that time there was a dominating idea that it is the Big Bang model complemented with the Inflation theory which is the adequate model of the Universe at least at first approximation. However it was in vain hope. The cosmological paradigm had to be changed under the force major of the observations made with ever increasing precision. Since Hubble's time the  cosmologists made attempts to measure the deceleration of the expansion caused by gravity. Confidence to discover namely the deceleration effect was so firm that the corresponding parameter was called the deceleration parameter (DP). However in 1998 two independent collaborations \cite{riess,perlmutter}, having researched distant supernovae, presented convincing evidence of the fact that the expansion of Universe is accelerated. It turned out that the brightness decreases in average considerably faster than it was commonly derived from the Big Bang model. Such additional dimming means that a given red shift corresponds to some additional distance. And it shows that the cosmological expansion is accelerated: the Universe expanded in the past slower than today. The discovery of the cosmic acceleration was probably one of the most importance not only for modern cosmology but also for physics in general. The Universe expanding in accelerated way represents the most direct demonstration of the fact that our fundamental theories are either incomplete or, even worse, incorrect \cite{silvestri_trodden,smolin}.

We would like to stress however that, in spite the great importance of the above mentioned discovery, the role of the DP in cosmological dynamics is far from being limited by that. The first of the two basic cosmological parameters (the Hubble parameter) describes the linear part of the time dependence of the scale factor (SF).  The linear evolution is dull and monotonous. All linear system behave mostly in the same way while every non-linear system is "unhappy in its own manner" (paraphrasing Lev Tolstoi). Linear dynamics can be only regular while non-linear one gives much more possibilities. Existence of bifurcation points in a non-linear system make dynamical regimes to change. Presence of local instability brings possibility to exist for chaotic regimes. Treating the Universe as a dynamical system it is natural to assume that it is non-linear: indeed, linearity is nothing more than approximation, while non-linearity represents the generic case. It is evident that future models of the Universe must take into account different aspects of its evolution. However the simplified model of homogeneous and isotropic Universe based on the cosmological principle contains single dynamical variable in form of the scale factor and description of its time evolution presents basic task of the theory. It is the first terms in the Taylor series for the scale factor $a(t)$ in the vicinity of the present time $t_0$ who are determined by the parameters $H_0$ (the linear part) and $q_0$ (the first non-linear correction).

As soon as the scale factor is the only dynamical variable, the quantities which determine its time dependence must be essentially present in all aspects of the Universe' evolution. Basic characteristics of the cosmological evolution, both static and dynamical, can be expressed in terms of the parameters $H_0$ and $q_0$. The very parameters (and higher time derivatives of the scale factor) enable us to construct model-independent kinematics of the cosmological expansion.

Time dependence of the scale factor reflects main events in history of the Universe. Moreover it is the DP who dictates the expansion rate of the Hubble sphere and determines the dynamics of the observable galaxy number variation: depending on the sign of the DP this number either grows (in the case of decelerated expansion), or we are going to stay absolutely alone in the cosmos (if the expansion is accelerated).

The intended purpose of the report is reflected in its title - "Cosmology in terms of the deceleration parameter". We would like to show that practically any aspect of the cosmological evolution is tightly bound to the DP. We try to make the report possibly independent on the current conjecture associated with the discovery of the accelerated expansion of the Universe. The formula $S=v_0t+at^2/2$ is correct regardless of the DP sign. On the other hand we cannot stay aside from the current observational status of the DP.

In the conclusions of the paper, by commenting about future observational values of $H_0$ and $q_0$, Sandage wrote: "The present discussion is only a prelude to the coming decade. If work now in progress is successful, better values for both $H_0$ and $q_0$ (and perhaps even $\Lambda$) should be found, and the 30-year dream of choosing between world models on the basis of kinematics alone might possibly be realized".

A quarter of a century then passed, a new generation of cosmologists came, but the world still was in waiting of the "correct" model of the Universe. A hope for happy outcome of the waiting was again associated to forthcoming determination of the DP.  In 1997 M. Turner wrote \cite{turner}: "At the moment, $\Lambda$CDM best accommodates all the observations, but I believe the evidence is not yet strong enough to abandon the other CDM models. Especially because additional observations will soon be able to decisively distinguish between the different models... This observation include DP $q_0$. LCDM predicts $q_0\sim1/2$, while the other CDM models predict $q_0=1/2$. Two groups (The Supernova Cosmology Project and The High $z$ Supernova Team) are hoping to determine $q_0$ to a precision of $\pm0.2$ by using distant Type Ia supernovae ($z\sim0.3-0.7$) as standard candles. Together, they discovered more than 40 high redshift supernovae last fall and winter and both groups should be announcing results soon."

Yet two decades almost passed. Both above mentioned groups were awarded by the Nobel Prize. However the DP continues to play central role both in description of cosmological observations and in theoretical discussions. And it is in spite of the appearance of the new cosmological favorite - the dark energy. As we have mentioned above, just a few decades before the cosmology could be defined as a discipline involved in study of only two numbers: current Universe' expansion rate $H_0$ and the DP $q_0$. Now we know these numbers with reasonable accuracy but it turns out that fate of the Universe remains still undefined. To predict the fate of the Universe one needs not only the current values of the parameters but their time dependence too if present. The accelerations are determined by forces and therefore to determine the Universe' fate one needs not only kinematics but dynamics of the cosmological expansion too. The latter is known to be determined by energy content of the Universe, the concept of which had drastically changed in the recent time. The exotic dark energy unexpectedly appeared (rather came back) on the scene and now it seemingly governs dynamics of the Universe. The dark energy (in form of the cosmological constant or in different dynamical forms generated by scalar fields) finally balanced the energy budget and made total density of the energy content in the Universe equal to the critical value predicted by the inflation theory. The new energy component has negative pressure and as a consequence it causes accelerated expansion of the Universe. It affects both the past evolution of the Universe and its future history. If the future for the dark energy in form of the cosmological constant is sufficiently clear and it represents monotonic accelerated expansion, then for the dynamical forms of the dark energy there is a whole ensemble of different possibilities: Big Rip, Big Whimper, Big Decay, Big Crunch, Big Brunch, Big Splat etc.

We stress one more time that the New Cosmology is based on the accelerated expansion of the Universe discovered at the end of the last century. Although the 2011 Nobel Prize award for the discovery of the accelerated expansion of the Universe formally transferred this effect (if this grandious phenomenon can be called simply "an effect") into the status of physical reality, summarizing the current state of affairs one must honestly admit that we now understand very little about the reason why the Universe' expansion is accelerated. The remaining doubts can be overcome by the cross-check of the expansion kinematics based on different physical mechanisms. If the fundamental conclusion on the accelerated expansion of the Universe turns out to be false then it is difficult to expect the consistency of the results obtained in different approaches (expect in the result of very unlikely random coincidence). While the most impressive aspect of the cosmological acceleration consists in the fact that the different strategies of its description lead to the same conclusion: the Universe entered into the stage of the accelerated expansion. However experience teaches us never relax our vigilance. A. Eddington is attributed the saying: do not trust the observations too much until they are not confirmed by theory. This paradoxical notion is not a theoretical arrogance, it reflects deeper understanding that not a plain collection of facts but their understanding makes the science. General Relativity allows the accelerated expansion of the Universe filled with a negative pressure substance, but it does not provide yet deeper understanding of this phenomenon. Further studies of the kinematics and dynamics of the Universe' expansion will provide long-time work for both the cosmologists-experimentalists (observers) and theorists.
\chapter{Background}
\section{Definitions and simplest relations}
The Hubble parameter and the DP represent simplest cosmographic parameters. It is well known \cite{weinberg,ellis} that the cosmological principle enables us to construct metrics of the Universe and make first steps towards interpretation of the cosmological observations. Recall that the kinematics is the part of mechanics responsible for description of a body's motion disregarding the forces which cause it. In the sense cosmography is just kinematics of cosmological expansion. In order to build the key characteristic -- time dependence of the scale factor $a(t)$ -- one needs to take equations of motion (the Einstein's equations) and make an assumption about material content of the Universe, which allow to construct the energy-momentum tensor. The cosmography is efficient because it allows to test any cosmological model which does not contradict the cosmological principle. Modifications of General Relativity or introduction of new components (such as dark matter or dark energy) evidently change the dependence $a(t)$ but it affects in no way the relations between the kinematic characteristics.

The parameters $H(t)$ and $q(t)$ can naturally be defined making use of Taylor series for the scale factor $a(t)$ in vicinity of the present time $t_0$:
\begin{equation}\label{a(t)_teylor}
    a(t)=a\left( {{t}_{0}} \right)+\dot{a}\left( {{t}_{0}} \right)\left[ t-{{t}_{0}} \right]+\frac{1}{2}\ddot{a}\left( {{t}_{0}} \right){{\left[ t-{{t}_{0}} \right]}^{2}}+\cdots
\end{equation}
This relation can be rewritten in the form:
\begin{equation}\label{a(t)_teylor1}
   \frac{a(t)}{a\left( t_0\right)}=1+H_0\left[ t-t_0\right]-\frac{q_0}{2}H_{0}^{2}{{\left[ t-{{t}_{0}} \right]}^{2}}+\cdots,
\end{equation}
Dynamics of Universe is described in frames of General relativity by the Einstein equations:
\[R_{\mu\nu}-\frac12Rg_{\mu\nu}=8\pi GT_{\mu\nu},\]
where the energy momentum tensor $T_{\mu\nu}$ describes spatial distribution of mass(energy) while components of the curvature tensor $R_{\mu\nu}$ and its trace $R$ are expressed in terms of metric tensor $g_{\mu\nu}$ and its derivatives of first and second order. The Einstein equations in general are non-linear and complicated to solve. The problem is simplified if one considers mass distribution with special symmetry properties embedded in the metrics. For homogeneous and isotropic Universe described by the Friedman-Lema\^{\i}tre-Robertson-Walker (FLRW) metrics \ref{flrw_metrics} the Einstein equations are reduced to system of two Friedman equations:
\begin{align}
\frac{\dot{a}^2}{a^2}=&\frac{1}{3M^2_{Pl}}\rho-\frac{k}{a^2},\label{background_2_5_1}\\
\frac{\ddot a}{a}=-&\frac{1}{6M^2_{Pl}}(\rho+3p),\label{background_2_5_2}
\end{align}
where $\rho$ and $p$ are respectively total density and pressure of all components present in the Universe at the considered moment of time and $M_{Pl}\equiv (8\pi G)^{-1/2}$ is the reduced Planck mass. This system of equations is incomplete: the two equations are not sufficient to completely describe the dynamics of Universe. Lorentz-invariance of the energy-momentum tensor $T_{\nu,\mu}^{\mu}=0$ leads to the conservation equation
\begin{equation}\label{conservation_equation}
\dot\rho+3H(\rho+p)=0.\end{equation}
Let us introduce relative density of i-th component of the energy density
\[\Omega_i\equiv\frac{\rho_i}{\rho_c};\quad \rho_c\equiv M^2_{Pl}H^2,\quad \Omega\equiv\sum\limits_i\Omega_i.\]
Having defined relative density for curvature \[\Omega_k\equiv-\frac{a^2}{H^2},\] one obtains the first Friedman equation in the following form
\[\sum\limits_i\Omega_i=1.\]

In order to solve the Friedman equations one needs to define the matter content in the Universe and construct the state equation for each component. In the simplest linear parametrization the state equation takes the form
\[p_i=w_i\rho_i\]
Solution of the Friedman equations for the case of one-component spatially flat $(k=0)$ Universe with time-independent state equation parameter $(w=w_i=const)$ reads
\begin{equation}\label{background_2_10}
a(t)\propto\left(\frac{t}{t_0}\right)^{\frac2{3(1+w)}},\quad \rho\propto a^{-3(1+w)}
\end{equation}
We normalize the scale factor by the condition $a(t_0)=1$. Such solutions exist only in the case $w\ne-1$. For the Universe dominated by radiation (relativistic gas of photons and neutrinos) one has $w=1/3$, while $w=0$ for the matter-dominated case. Then (\ref{background_2_10}) takes the form
\[a(t)\propto\left(\frac{t}{t_0}\right)^{2/3},\quad \rho\propto a^{-3}\]
for matter and
\[a(t)\propto\left(\frac{t}{t_0}\right)^{1/2},\quad \rho\propto a^{-4}\]
for radiation.

The condition $\rho=const$ (as for cosmological constant) requires $w=-1$. In such a case the Hubble speed remains constant leading to exponential growth of the scale factor
\[a(t)\propto e^{Ht}.\]
Therefore both traditional cosmological components -- matter $(w=0)$ and radiation $(w=1/3)$ -- can produce only decelerated expansion of Universe $(\ddot a<0,\ q>0)$, namely
\begin{equation}\label{background_2_14}
q=\left\{
\begin{array}{ll}
1 & {\rm for\ radiation} \\
1/2 & {\rm for\ matter}\\
-1 & {\rm for\ cosmological\ constant}
\end{array}
\right.
\end{equation}
Using the definition of the DP one obtains for the case of spatially flat Universe filled with single component with the state equation $(p=w\rho)$ the following result
\begin{equation}\label{background_2_15}q=\frac12(1+3w).\end{equation}
In general,
\[k=0,\pm1,\quad \rho=\sum\limits_i\rho_i,\quad p=\sum\limits_i\rho_i w_i\]
and one obtains
\[q=\frac\Omega2+\frac32\sum\limits_i w_i \Omega_i,\]
or
\[q=\frac12\frac{H_0^2}{H^2}\sum\limits_i \Omega_{i0}(1+z)^{3(1+w_i)}(1+3w_i).\]
The latter equation can be rewritten in the form
\[q=\frac12\left(1+\frac{k}{a^2H^2}\right)\left(1+3\frac{p}{\rho}\right)\]
or equivalently
\[q=\frac12\left[1+3\sum\limits_i w_i \Omega_i\right]+\frac{k}{2a^2H^2}.\]

Let us consider in more detail the model of two-component Universe \cite{ponce_de_leon}. Such approximation is sufficient to achieve good accuracy on each stage of its evolution. At the present time the two dark components -- dark matter and dark energy -- are considered to be dominating. With this in mind, let us represent total density and pressure in the form
\begin{align}
\nonumber \rho & = \rho_{de} +\rho_{m};\\
\nonumber p & = p_{de} +p_{m}.
\end{align}
We neglect meanwhile the interaction between the components and as a result they separately satisfy the conservation equation (\ref{conservation_equation}). Let us assume that the state equation parameter for each component depends on the scale factor
\begin{align}
\nonumber p_{de} & = W(a) \rho_{de};\\
\nonumber p_m & = w(a)\rho_{m}.
\end{align}
With these assumptions
\[\Omega_m +\Omega_{de}=1+\frac{k}{a^2H^2},\]
\[q=\frac12+\frac32\left[W\Omega_{de}+w\Omega_m\right]+\frac{k}{2a^2H^2}.\]
These can, formally, be regarded as two equations for $\Omega_{de}$ and $\Omega_{m}$. Solving them we get
\begin{align}
\nonumber \Omega_{m} & = \frac{2q-1-3W}{3(w-W)}-\frac{k(1+3W)}{3a^2H^2(w-W)};\\
\nonumber \Omega_{m} & = \frac{2q-1-3w}{3(w-W)}-\frac{k(1+3w)}{3a^2H^2(w-W)}.
\end{align}
We note that the denominator in these expressions is always positive because $W<0$ for dark energy. Thus, the fact that $\Omega_{m(de)}>0$ imposes an upper and lower limit on $q$,
\[\frac{1+3W}{2}\left(1+\frac{k}{a^2H^2}\right)\le q\le \frac{1+3w}{2}\left(1+\frac{k}{a^2H^2}\right).\]
It is easy to establish relation between the DP and the pressure for the case of spatially flat one-component. It follows from the conservation equation (\ref{conservation_equation}) that
\[p=-\frac{\dot\rho}{3H}\frac{w}{1+w}.\]
Using
\[w = \frac{2q-1}3;\quad \dot\rho=\frac{3}{4\pi G}H\dot H,\quad \dot H=-H^2(1+q)\]
one finds
\[p=\frac{H^2}{8\pi G}(2q-1).\]
For spatially flat one-component Universe with state equation $p=w\rho$ the scale factor is \[a\propto t^\frac{2}{3(1+w)}\] and therefore \[H=\frac{2}{3(1+w)}\frac1t.\]

Using (\ref{background_2_15}) one can find a simple relation between the current age of the Universe and the DP
\begin{equation}\label{background_2_22}t_0=\frac{H_0^{-1}}{1+q}.\end{equation}

Let us now give an instructive example of the fact that solution of the Friedman equations can be naturally  represented in terms of the two parameters $H_0$ and $q_0$ for the case of Universe filled with non-relativistic matter \cite{shtanov}. Let us start with closed Universe ($k=+1$). The Friedman equations can then be represented in the form
\begin{align}
\label{background_2_29} H^2 & +\frac1{a^2} =\frac{8\pi G}3\rho,\\
\nonumber 2\frac{\ddot a}{a} & +H^2+\frac1{a^2} = 0.
\end{align}
Using this equations one can find the relations between the current values of the Universe's parameters
\begin{align}
\label{background_2_30} H_0^2 & =\frac1{a_0^2(2q_0-1)},\\
\nonumber q_0 & = \frac{4\pi G}{3H_0^2}\rho_0.
\end{align}
Note that in general
\[q_0=\frac{4\pi G}3\frac{\rho_0+3p_0}{H_0^2}=\frac{\rho_0+3p_0}{2\rho_{0,crit}},\quad \rho_{0,crit}=\frac{3H_0^2}{8\pi G}.\]
For Universe filled only with non-relativistic matter one has $\Omega_m = 2q_0$. It is easy to see that $q_0>1/2$ and $\Omega_m>1$ as was expected in the closed model. Using (\ref{background_2_30}) one can rewrite the equation for scale factor in the form
\begin{equation}\label{background_2_31}\dot a^2=\frac\alpha a-1,\quad \alpha\equiv\frac{2q_0}{H_0(2q_0-1)^{3/2}}.\end{equation}
It is easy to see that the considered model includes both parameters $H_0$ and $q_0$. integration of (\ref{background_2_31}) gives \[t=\int \sqrt{\frac{a}{\alpha-a}}\,da.\]
Substitution \[a=\frac\alpha2(1-\cos\tau)\] leads to \begin{equation}\label{background_2_33}t=\frac\alpha2(\tau-\sin\tau).\end{equation}
Because of the relation $dt=ad\tau$ it is evident that the variable $\tau$ is the conformal time. Taking the constants of integration so that $a=0$ as $t=0$ (and $\tau=0$) we can see that $a=a_0$ at $\tau=\tau_0$. Consequently, \[\cos\tau_0=\frac{1-q_0}{q_0},\quad \sin\tau_0=\frac{\sqrt{2q_0-1}}{q_0}.\]
From (\ref{background_2_33}) it follows that the age of Universe in close model is \[t_0=\frac\alpha2(\tau_0-\sin\tau_0)=\frac{q_0}{H_0(2q_0-1)^{3/2}}\left(\arccos{\frac{1-q_0}{q_0}} - \frac{\sqrt{2q_0-1}}{q_0}\right).\]
The maximum of scale factor reaches at $\tau=\pi$,
\[a_{\rm max}=\alpha=\frac{2q_0}{H_0(2q_0-1)^{3/2}}.\]
As one can see, all parameters of the model can be expressed in terms of parameters $H_0$ and $q_0$.

In the case of open Universe the formulae (\ref{background_2_29})-(\ref{background_2_31}) transform into
\begin{align}
\label{background_2_37} H^2 & -\frac1{a^2} =\frac{8\pi G}3\rho,\\
\nonumber 2\frac{\ddot a}{a} & +H^2-\frac1{a^2} = 0,\\
\label{background_2_38} H_0^2 & =\frac1{a_0^2(1-2q_0)},\\
\nonumber q_0 & = \frac{4\pi G}{3H_0^2}\rho_0,
\end{align}
\begin{equation}\label{background_2_39}\dot a^2=\frac\beta a+1,\quad \beta\equiv\frac{2q_0}{H_0(1-2q_0)^{3/2}}.\end{equation}
Now $q_0<1/2$, $0\le\Omega_m< 1$. The solution of (\ref{background_2_39}) in the conformal time parametrization is \[a=\frac\beta2(\cosh\tau-1),\quad t=\frac\beta2(\sinh\tau-\tau).\] Current value $\tau_0$ of the conformal time is defined by the relation \[\cosh\tau_0=\frac{1-q_0}{q_0}.\]
The age of the Universe is
\[t_0=\frac\beta2(\sinh\tau_0-\tau_0) = \frac{q_0}{H_0(1-q_0)^{3/2}}\left(\frac{\sqrt{1-2q_0}}{q_0} - \ln\frac{1-q_0+\sqrt{1-2q_0}}{q_0}\right).\]
In this case again all characteristics of the model are expressed in terms of the parameters $H_0$ and $q_0$.

\section{Classification of models of Universe based on the deceleration parameter}
When the rate of expansion never changes, and $\dot a$ is constant, the scaling factor is proportional to time $t$, and the deceleration term is zero. When the Hubble term is constant, the deceleration term $q$ is also constant and equal to $-1$, as in the de Sitter and steady-state Universes. In most Universes the deceleration term changes in time. One can classify models of Universe on the basis of time dependence of the two parameters. All models can be characterized by whether they expand or contract, and accelerate or decelerate:
\begin{description}
  \item[(a)]$H>0$, $q>0$: expanding and decelerating
  \item[(b)] $H>0$, $q<0$: expanding and accelerating
  \item[(c)] $H<0$, $q>0$: contracting and decelerating
  \item[(d)]$H<0$, $q<0$: contracting and accelerating
  \item[(e)] $H>0$, $q=0$: expanding, zero deceleration
  \item[(f)] $H<0$, $q=0$: contracting, zero deceleration
  \item[(g)]$H=0$, $q=0$: static
\end{description}

Of course, generally speaking, both the Hubble parameter and DP can change their sign during the evolution. Therefore the evolving Universe can transit from one type to another. It is one of the basic tasks of cosmology to follow this evolution and clarify its causes. There is little doubt that we live in an expanding Universe, and hence only (a), (b), and (e) are possible candidates. Evidences in favor of the fact that the expansion is presently accelerating continuously grows in number and therefore the current dynamics belongs to type (b). Having fixed the material content we can classify the types using the connection between  the DP and the spatial geometry. For example, in the case of Universe filled with non-relativistic matter one gets
\begin{align}
\nonumber q>\frac12 & :\quad  closed\ spherical\ space & k&=+1;\\
\nonumber q=\frac12 & :\quad  open\ flat\ space & k&=0;\\
\nonumber q<\frac12 & :\quad  open\ hyperbolic\  space & k&=-1.
\end{align}

The classification can be also based on the connection between the DP and age of the Universe. Foe example, for the Einstein-de Sitter Universe where the age is \[t^*=\frac23H^{-1}\] we  have
\begin{align}
\nonumber q>\frac12 & :\quad  age<t^*;\\
\nonumber q=\frac12 & :\quad  age=t^*;\\
\nonumber q<\frac12 & :\quad  age>t^*.
\end{align}

Note that sign of the DP determines the difference between the actual age of the Universe and the Hubble time. In a decelerating Universe with $q>0$, the age of the Universe will be less than the Hubble time, because at earlier times it was expanding at a faster rate, whereas a Universe that has always been accelerating, that is, $q<0$ for all time, will have an age that is greater than the Hubble time. A Universe that expands at a constant rate, $q=0$, has an age equal to the Hubble time.

Finally, if we are interested solely in the expansion regime then in the case of constant DP the Universe would exhibit decelerating expansion if $q>0$, an expansion with constant rate if $q=0$, accelerating power-law expansion if $-1<q<0$, exponential expansion (also known as de Sitter expansion) if $q=-1$ and super-exponential expansion if $q<-1$.
\section{Deceleration Parameter in the $n$-dimensional Universe}
Following \cite{NOO}, consider an $(n+1)$-dimensional  homogeneous and isotropic
Lorentzian
spacetime with the metric
\begin{equation} \label{1}
ds^2=g_{\mu\nu} dx^\mu dx^\nu=-dt^2+a^2(t)g_{ij} dx^i dx^j,\quad i,j=1,\dots,n,
\end{equation}
where $t$ is the cosmological (or cosmic) time and $g_{ij}$ is the  metric of an $n$-dimensional Riemannian manifold
$M$ of constant scalar curvature characterized by an indicator, $k=-1,0,1$,
so that $M$ is an $n$-hyperboloid, the flat space $\Bbb R^n$, or an $n$-sphere, with the
respective metric
\begin{equation} \label{2}
g_{ij}dx^i dx^j=\frac1{1-kr^2}\,dr^2+r^2\,d\Omega^2_{n-1},
\end{equation}
where $r>0$ is the radial variable and $d\Omega_{n-1}^2$ denotes the canonical metric of the unit
sphere $S^{n-1}$ in $\Bbb R^n$. Inserting the metric (\ref{1})--(\ref{2}) into the Einstein equations
\begin{equation}
G_{\mu\nu}+\Lambda g_{\mu\nu}=8\pi G T_{\mu\nu},
\end{equation}
where $G_{\mu\nu}$ is the Einstein tensor, $G$ the universal gravitational constant, and $\Lambda$ the cosmological constant, the speed of light is set to unity, and $T_{\mu\nu}$ is the
energy-momentum tensor of an ideal cosmological fluid given by
\begin{equation} \label{4}
T^{\mu\nu}=\mbox{\footnotesize    diag}\{\rho_m,p_m,\dots,p_m\},
\end{equation}
with $\rho_m$ and $p_m$ the $t$-dependent matter energy density and pressure,
we arrive
at the Friedmann equations
\begin{align}
H^2=&\frac{16\pi G}{n(n-1)}\rho-\frac k{a^2},\label{5}\\
\dot{H}=&-\frac{8\pi G}{n-1}(\rho+p)+\frac k{a^2},\label{6}
\end{align}
in which $\rho,p$ are the effective energy
density and pressure related to $\rho_m,p_m$ through
\begin{equation} \label{8}
\rho=\rho_m+\frac{\Lambda}{8\pi G},\quad p=p_m-\frac{\Lambda}{8\pi G}.
\end{equation}
On the other hand, recall that, with (\ref{1}) and (\ref{4}) and (\ref{8}), the energy-conservation law, $\nabla_\nu T^{\mu\nu}=0$, takes the form
\begin{equation} \label{9}
\dot{\rho}_m+n(\rho_m+p_m)H=0.
\end{equation}
To proceed further, assume that the perfect-fluid cosmological model satisfies the
barotropic equation of state
\begin{equation} \label{10}
p_m=w \rho_m.
\end{equation}
Inserting (\ref{10}) into (\ref{9}), we have
\begin{equation}\label{11}
\dot{\rho}_m+n(1+w)\rho_m \frac{\dot{a}}a=0,
\end{equation}
which can be integrated to yield
\begin{equation}\label{12}
\rho_m=\rho_0 a^{-n(1+w)},
\end{equation}
where $\rho_0>0$ is an integration constant \cite{KL}. Using (\ref{12}) in (\ref{8}), we arrive at
the relation \cite{NOO,KL}
\begin{equation}\label{13}
\rho=\rho_0 a^{-n(1+w)}+\frac{\Lambda}{8\pi G}.
\end{equation}
From (\ref{5}) and (\ref{13}), we get the following equation of motion \cite{gibbons} for the scale factor $a$:
\begin{equation} \label{14}
\dot{a}^2=\frac{16\pi G\rho_0}{n(n-1)}a^{-n(1+w)+2}+\frac{2\Lambda}{n(n-1)} a^2-k.
\end{equation}

To integrate (\ref{14}), we recall Chebyshev's  theorem
\cite{C0,C1}: For rational numbers
$p,q,r$ ($r\neq0$) and nonzero real numbers $\alpha,\beta$, the integral \[\int x^p(\alpha+\beta x^r)^q\,dx\] is elementary
if and only if at least one of the quantities
\begin{equation}\label{cd}
\frac{p+1}r,\quad q,\quad \frac{p+1}r+q,
\end{equation}
is an integer.\footnote{Another way to see the validity of the Chebyshev theorem is to
represent the integral of concern by a hypergeometric function such that when a quantity in (\ref{cd}) is an integer the hypergeometric function
is reduced into an elementary function.}
 Consequently, when $k=0$ or $\Lambda=0$, and $w$ is rational, the Chebyshev theorem enables us to know that,
for exactly
what values of $n$ and $w$, the equation (\ref{14}) may be integrated.

We focus on the spatially flat situation $k=0$, which is known to be
most relevant  for cosmology \cite{nature,science,Copeland,DS},
and rewrite equation (\ref{14}) as
\begin{equation} \label{15}
\dot{a}=\pm\sqrt{c_0 a^{-n(1+w)+2}+\Lambda_0 a^2},\quad c_0=\frac{16\pi G\rho_0}{n(n-1)},\quad\Lambda_0=\frac{2\Lambda}{n(n-1)}.
\end{equation}
Then (\ref{15}) reads
\begin{equation} \label{15a}
\pm\int a^{-1}\left(c_0 a^{-n(1+w)}+\Lambda_0 \right)^{-\frac12}da=t+C.
\end{equation}
It is clear that the integral on the left-hand side of (\ref{15a}) satisfies the integrability condition stated in the Chebyshev theorem for any $n$ and any rational $w$.

It turns out
that (\ref{15}) might be integrated for any real $w$ as well,
not necessarily rational.
To do so we apply $a>0$ and get from (\ref{15}) the equation
\begin{equation} \label{16}
\frac{d}{dt}\ln a=\pm\sqrt{c_0 a^{-n(1+w)}+\Lambda_0},
\end{equation}
or equivalently,
\begin{equation}\label{17}
\dot{u}=\pm\sqrt{c_0 e^{-n(1+w)u}+\Lambda_0},\quad u=\ln a.
\end{equation}
Set
\begin{equation}\label{18}
\sqrt{c_0 e^{-n(1+w)u}+\Lambda_0}=v.
\end{equation}
Then
\begin{equation}\label{19}
u=\frac{\ln c_0}{n(1+w)}-\frac 1{n(1+w)} \ln(v^2-\Lambda_0).
\end{equation}
Inserting (\ref{19}) into (\ref{17}), we find
\begin{equation}\label{20}
\dot{v}=\mp\frac12 n(1+w)(v^2-\Lambda_0),
\end{equation}
whose integration gives rise to the expressions
\begin{equation} \label{21}
v(t)=\left\{\begin{array}{rl} &v_0\left(1\pm \frac12 n(1+w)v_0t\right)^{-1},\quad \Lambda_0=0;\\
&\\
& \sqrt{\Lambda_0}(1+C_0 e^{\mp n(1+w)\sqrt{\Lambda_0} t})(1-C_0 e^{\mp n(1+w)\sqrt{\Lambda_0} t})^{-1},\\& C_0=(v_0-\sqrt{\Lambda_0})(v_0+\sqrt{\Lambda_0})^{-1},\quad \Lambda_0>0;\\
&\\
&\sqrt{-\Lambda_0}\tan\left(\mp\frac12 n(1+w)\sqrt{-\Lambda_0} t +\arctan\frac{v_0}{\sqrt{-\Lambda_0}}\right),\quad
\Lambda_0<0,
\end{array}
\right.
\end{equation}
where $v_0=v(0)$. Hence, in terms of $v$, we obtain the time-dependence of the scale factor $a$:
\begin{equation} \label{22}
a^{n(1+w)}(t)=\frac{8\pi G \rho_0}{\frac12 n(n-1)v^2(t)-\Lambda}.
\end{equation}

We now assume
\(w>-1\) in the equation of state in our subsequent discussion.
We are interested in solutions satisfying
\(a(0)=0.\)

When $\Lambda=0$, we combine (\ref{21}) and (\ref{22}) to get
\begin{equation} \label{x1}
a^{n(1+w)}(t)=4\pi G\rho_0\left(\frac n{n-1}\right)(1+w)^2 t^2.
\end{equation}
When $\Lambda>0$, we similarly obtain
\begin{equation}\label{x2}
a^{n(1+w)}(t)=\frac{8\pi G\rho_0}{\Lambda}\sinh^2\left(\sqrt{\frac{n\Lambda}{2(n-1)}}(1+w) t\right).
\end{equation}
When $\Lambda<0$ we rewrite (\ref{22}) as
\begin{equation} \label{23}
a^{n(1+w)}(t)=\frac{8\pi G \rho_0}{(-\Lambda)}\cos^2\left(\sqrt{\frac{n(-\Lambda)}{2(n-1)} }(1+w)t \mp\arctan\sqrt{\frac{n(n-1)}{-2\Lambda}}\, v_0\right).
\end{equation}

If we require $a(0)=0$, then (\ref{23}) leads to the conclusion
\begin{equation} \label{24}
a^{n(1+w)}(t)=\frac{8\pi G \rho_0}{(-\Lambda)}\sin^2\sqrt{\frac{n(-\Lambda)}{2(n-1)} }(1+w)t,
\end{equation}
which gives rise to a periodic Universe so that the scale factor $a$ reaches its maximum $a_m$,
\begin{equation} \label{25}
a^{n(1+w)}_m=\frac{8\pi G \rho_0}{(-\Lambda)},
\end{equation}
at the times
\begin{equation}\label{26}
t=t_{m,k}=\left(\frac\pi2+k\pi\right)\frac1{(1+w)}\sqrt{\frac{2(n-1)}{n(-\Lambda)}},\quad k\in\Bbb Z,
\end{equation}
and shrinks to zero at the times
\begin{equation}\label{27}
t=t_{0,k}=\frac{k\pi}{(1+w)}\sqrt{\frac{2(n-1)}{n(-\Lambda)}},\quad k\in\Bbb Z.
\end{equation}

Using the equations (\ref{x1},\ref{x2},\ref{24}), one easily obtains the expressions for the DP
\begin{equation}
q(t)=\left\{\begin{array}{cc} \frac n2 (1+w)-1, & \Lambda_0=0;\\
&\\
\frac{n(1+w)}{2\cosh^2\left(\sqrt{\frac{n\Lambda}{2(n-1)}}(1+w)t\right)}-1, & \Lambda_0>0;\\
&\\
\frac{n(1+w)}{2\cos^2\left(\sqrt{\frac{n\Lambda}{2(n-1)}}(1+w)t\right)}-1, &
\Lambda_0<0.
\end{array}
\right.
\end{equation}

\section{Deceleration as a cosmographic parameter}
In the present section we will use an approach to describe the Universe which is called a ``cosmography'' \cite{weinberg1}. It is entirely based on the cosmological principle, stating that the Universe is homogeneous and isotropic on scales larger than a hundred megaparsecs. It allows to choose among whole possible variety of models describing the Universe a narrow set of homogeneous and isotropic models. The most general space-time metrics agreed with the cosmological principle is the FLRW one
\begin{equation}\label{flrw_metrics}
ds^2 = dt^2-a^2(t)\left\{\frac{dr^2}{1-kr^2}+r^2(d\theta^2+\sin^2\theta d\varphi^2)\right\}.
\end{equation}
Here $r,\theta,\varphi$ are coordinates of the point which does not participate in any motion except the global expansion of the Universe, that is they are the comoving coordinates.

The cosmological principle enables us to build the metrics of the Universe and make first steps towards interpretation of the cosmological observations. Recall that the kinematics is the part of mechanics responsible for description of a body's motion disregarding the forces which cause it. In the sense cosmography is just kinematics of cosmological expansion. In order to build the key characteristic -- time dependence of the scale factor $a(t)$ -- one needs to take equations of motion (the Einstein's equations) and make an assumption about material content of the Universe, which allow to construct the energy-momentum tensor. The cosmography is efficient because it allows to test any cosmological model which does not contradict the cosmological principle. Modifications of General Relativity or introduction of new components (such as dark matter or dark energy) evidently change the dependence $a(t)$ but it affects in no way the relations between the kinematic characteristics.

Expansion rate of the Universe is defined by the Hubble parameter $H\equiv\dot a(t)/a(t)$ which depends on time. The time dependence is measured by the DP $q(t)$. In order to make more detailed description of kinematics of cosmological expansion it is useful to consider the extended set of the parameters which includes higher order time derivatives of the scale factor \cite{visser1,visser2,capozziello}:
\begin{align}
\nonumber H(t)&\equiv\frac1a\frac{da}{dt};\\
\nonumber q(t)&\equiv-\frac1a\frac{d^2a}{dt^2}\left[\frac1a\frac{da}{dt}\right]^{-2};\\
\nonumber j(t)&\equiv\frac1a\frac{d^3a}{dt^3}\left[\frac1a\frac{da}{dt}\right]^{-3};\\
\nonumber s(t)&\equiv\frac1a\frac{d^4a}{dt^4}\left[\frac1a\frac{da}{dt}\right]^{-4};\\
\nonumber l(t)&\equiv\frac1a\frac{d^5a}{dt^5}\left[\frac1a\frac{da}{dt}\right]^{-5}.
\end{align}
Note that the four latter parameters are dimensionless. In terms of the conformal time \[H=\frac{a'}{a^2},\quad q=-\left(\frac{aa''}{a'^2}-1\right),\]
where prime denotes derivative  w.r.t. the conformal time.

We will concentrate our attention on the DP $q(t)$. However one should note unique importance of the jerk parameter $j(t)$, in particular for testing of the cosmological models. "It is a striking and slightly puzzling fact that almost all current cosmological observations can be summarized by the simple statement: The jerk of the Universe equals one" \cite{dunajski_gibbons}.

Let us make Taylor expansion of the scale factor in time using the above introduced parameters
\begin{align}
\nonumber a(t)=&a_0\left[1+H_0(t-t_0)-\frac12q_0H_0^2(t-t_0)^2+ \frac1{3!}j_0H_0^3(t-t_0)^3\right.\\
\label{background_4_3} &\left.+\frac1{4!}s_0H_0^4(t-t_0)^4+\frac1{5!}l_0H_0^5(t-t_0)^5 + O\left((t-t_0)^6\right)\right].
\end{align}
Using the same parameters the Taylor expansion of the redshift takes on the form
\begin{align}
\nonumber 1+z=&\left[1+H_0(t-t_0)-\frac12q_0H_0^2(t-t_0)^2+ \frac1{3!}j_0H_0^3(t-t_0)^3\right.\\
\nonumber &\left.+\frac1{4!}s_0H_0^4(t-t_0)^4+\frac1{5!}l_0H_0^5(t-t_0)^5 + O\left((t-t_0)^6\right)\right]^{-1}.\\
\nonumber z=&H_0(t-t_0)+\left(1+\frac{q_0}2\right)H_0^2(t-t_0)^2+\cdots.
\end{align}

Derivatives of the lower order parameters can be expressed through the higher ones, for instance
\[\frac{dq}{d\ln(1+z)}=j-q(2q+1).\]
Let us give a few useful relations for the DP
\begin{align}
\label{background_4_6} q(t)=&\frac{d}{dt}\left(\frac1H\right)-1;\\
\nonumber q(z)=&\frac12\frac{d\ln H^2}{d\ln(1+z)}-1;\\
\nonumber q(z)=&\frac{d\ln H}{dz}(1+z)-1;\\
\nonumber q(a)=&-\left(1+\frac{\frac{dH}{dt}}{H^2}\right)=-\left(1+\frac{a\frac{dH}{da}}{H}\right)-1;\\
\nonumber q(a)=&\frac{d\ln(aH)}{d\ln a};\\
\nonumber q=&-1+(1+z)E^{-1}\frac{dE}{dz},\quad E\equiv\frac{H}{H_0};\\
\nonumber E=&e^{I(z)},\quad I(z)=\int\limits_0^z\frac{1+q(x)}{1+x}dx.
\end{align}

For a single-component fluid with density $\rho$ \[q(a)=-1-\frac{a\frac{d\rho}{da}}{2\rho}.\]
The derivatives $dH/dz$, $d^2H/dz^2$, $d^3H/dz^3$ and $d^4H/dz^4$ can be expressed through the DP $q$ and other cosmological parameters
\begin{align}
\label{background_4_8} \frac{dH}{dz}=&\frac{1+q}{1+z}H;\\
\nonumber \frac{d^2H}{dz^2}=&\frac{j-q^2}{(1+z)^2}H;\\
\nonumber \frac{d^3H}{dz^3}=&\frac{H}{(1+z)^3}\left(3q^2+3q^3-4qj-3j-s\right);\\
\nonumber \frac{d^4H}{dz^4}=&\frac{H}{(1+z)^4}\left(-12q^2-24q^3-15q^4+32qj+25q^2j+7qs+12j-4j^2+8s+l\right).
\end{align}
Formulae (\ref{background_4_8}) allow make series expansion of the Hubble parameter in terms of the scale factor
\[H(z)=H_0+\left.\frac{dH}{dz}\right|_{z=0}z +\frac12\left.\frac{d^2H}{dz^2}\right|_{z=0}z^2 +\frac16\left.\frac{d^3H}{dz^3}\right|_{z=0}z^3+\dots=\]
\[=H_0(1+(1+q_0)z+\frac12(j_0-q_0^2)z^2+\frac16(3q_0^2+3q_0^3-4q_0j_0-3j_0-s)z^3).\]
There is a decomposition for the inverse Hubble parameter
\begin{align}
\nonumber \frac{d}{dz}\left(\frac1H\right)=&-\frac1{H^2}\frac{dH}{dz}=-\frac{1+q}{1+z}\frac1H;\\
\nonumber \frac{d^2}{dz^2}\left(\frac1H\right)=& 2\left(\frac{1+q}{1+z}\right)^2\frac1H =\left(\frac{2+4q+3q^2-j}{(1+z)^2}\right)\frac1H ;\\
\nonumber \frac1{H(z)}=& \frac1{H_0}\left[1-(1+q_0)z+\frac{2+4q_0+3q_0^2-j_0}{6}z^2+\dots\right].
\end{align}

Let us cite some useful relations which allow to transform from higher order time derivatives to derivatives w.r.t. red shift:
\begin{align}
\nonumber \frac{d^2}{dt^2}=&(1+z)H\left[H+(1+z)\frac{dH}{dz}\right]\frac{d}{dz}+(1+z)^2H^2\frac{d^2}{dz^2},\\
\nonumber \frac{d^3}{dt^3}=&-(1+z)H\left\{H^2 +(1+z)^2\left(\frac{dH}{dz}\right)^2+ (1+z)H\left[4\frac{dH}{dz}+(1+z)\frac{d^2H}{dz^2}\right]\right\}\frac{d}{dz}\\
\nonumber & -3(1+z)^2H^2\left[H+(1+z)\frac{dH}{dz}\right]\frac{d^2}{dz^2}-(1+z)^3H^3\frac{d^3}{dz^3},\\
\nonumber \frac{d^4}{dt^4}=&(1+z)H\left[H^2 +11(1+z)H^2\frac{dH}{dz} +11(1+z)H\frac{dH}{dz} +(1+z)^3\left(\frac{dH}{dz}\right)^3\right.\\
\nonumber & \left. +7(1+z)^2H\frac{d^2H}{dz^2} +4(1+z)^3\frac{dH}{dz}\frac{d^2H}{dz^2} +(1+z)^3H^2\frac{d^3H}{dz^3}\right]\frac{d}{dz}\\
\nonumber &+(1+z)^2H^2\left[7H^2 +22H\frac{dH}{dz} +7(1+z)^2H\left(\frac{dH}{dz}\right)^2 +4H\frac{d^2H}{dz^2}\right]\frac{d^2}{dz^2}\\
\nonumber & +6(1+z)^3H^3\left[H+(1+z)\frac{dH}{dz}\right]\frac{d^3}{dz^3} +(1+z)^4H^4\frac{d^4}{dz^4}.
\end{align}
Derivatives of the Hubble parameter squared w.r.t. the redshift $d^iH^2/dz^i$, $i=1,2,3,4$ in terms of the cosmographic parameters take on the form
\begin{align}
\nonumber \frac{dH^2}{dz}=&\frac{2H^2}{1+z}(1+q),\\
\nonumber \frac{d^2H^2}{dz^2}=&\frac{2H^2}{(1+z)^2}(1+2q+j),\\
\nonumber \frac{d^3H^2}{dz^3}=&\frac{2H^2}{(1+z)^3}(-qj-s),\\
\nonumber \frac{d^4H^2}{dz^4}=&\frac{2H^2}{(1+z)^4}(4qj+3qs+3q^2j-j^2+4s+l).
\end{align}
The current values of deceleration and jerk parameters in terms of $N\equiv-\ln(1+z)$ read
\begin{align}
\nonumber q_0=& \left.-\frac1{H^2}\left\{\frac12\frac{d\left(H^2\right)}{dN}+H^2\right\}\right|_{N=0},\\
\nonumber j_0=&\left.\left\{\frac1{2H^2}\frac{d^2\left(H^2\right)}{dN^2} +\frac3{2H^2}\frac{d\left(H^2\right)}{dN}+1\right\}\right|_{N=0}.
\end{align}
Time derivatives of the Hubble parameter can be also expressed in terms of the cosmographic parameters
\begin{align}
\label{background_4_14} \dot H=& -H^2(1+q),\\
\nonumber \ddot H=& H^3(j+3q+2),\\
\nonumber \dddot H=& H^4[s-4j-3q(q+4)-6],\\
\nonumber \ddddot H=& H^5[l-5s+10(q+2)j+30(q+2)q+24].
\end{align}

It is easy to see from (\ref{background_4_14}) that accelerating growth of the expansion rate $\dot H>0$ corresponds to $q<-1$.

At last, from the power series expansion of the scalar factor (\ref{background_4_6}) one can also express the DP as a power series in time. This time variable can be written as a power series in redshift $z$ or $y$-redshift $z/(1+z)$, yielding respectively \cite{guimaraes_lima}
\begin{align}
\nonumber q(z)=&q_0 + \left(-q_0-2q_0^2+j_0\right)z+\frac12\left(2q_0+8q_0^2+8q_0^3-7q_0j_0-4j_0-s_0\right)z^2+O(z^3),\\
\nonumber q(y)=&q_0 +\left(-q_0-2q_0^2+j_0\right)y+\frac12\left(4q_0+8q_0^3-7q_0j_0-2j_0-s_0\right)y^2+O(y^3).
\end{align}
As can be seen from the relations (\ref{background_4_6}), the Hubble parameter is connected to the DP by the integral relation
\[H=H_0\exp\left[\int\limits_0^z[q(z')+1]d\ln (1+z')\right].\]
Then it immediately follows that one needs the information on the dynamics of cosmological expansion coded in the quantity $q(z)$ in order to reconstruct the basic characteristic of the expanding Universe which is $H(z)$.
\subsection{Cosmological scalars and the Friedmann equation}
Dunajski and  Gibbons \cite{dunajski_gibbons} proposed an original way to test the General Relativity (GR) and the cosmological models based on it. The procedure implies expressing the Friedmann equation in terms of directly measurable cosmological scalars constructed out of higher derivatives of the scale factor, i.e. cosmographic parameters $H,q,j,s,l$. In other words, the key idea is to treat the Friedmann equations as one algebraic constraint between the scalars. This links the measurement of the cosmological parameters to a test of GR, or any of its modifications (which would lead to different constraints).
For an example consider a Universe containing the cosmological constant $\Lambda$ and non-relativistic matter (dust). The Einstein equations reduce to the Friedmann equation
\begin{equation}\label{scalar_1}\dot a^2+k=\frac13\rho a^2+\frac13\Lambda a^2\end{equation}
and the conservation equation, which can be used to find the dependence of density on the scale factor ($\rho a^3=M$, $M=const$). Let us now consider a system of three equations consisting of (\ref{scalar_1}) and its first two time derivatives. We regard this as a system of algebraic equations for the constants ($k,\Lambda,M$) which can therefore be expressed as functions of $a,\dot a,\ddot a,\dddot a$. Take the third derivative of (\ref{scalar_1}) and substitute the expressions for ($k,\Lambda,M$).  The resulting equation does not contain any parameters and can be expressed in terms of the cosmological scalars as
\begin{equation}\label{scalar_2}s+2(q+j)+qj=0.\end{equation}
This fourth order ODE is equivalent to the Friedmann equation and has an advantage that it  appears as a constraint on directly measurable quantities. Thus it provides the test of the model basing on which the latter equation was obtained.

If only two constants ($\Lambda,M$) are eliminated between (\ref{scalar_1}) and its first derivative then the second derivative of (\ref{scalar_1}) yields \cite{harrison_1976}
\begin{equation}\label{scalar_3}k=a^2H^2(j-1).\end{equation}
where $k$ is regarded as a parameter. In particular if $k=0$ (SCM)  this relation reduces to a third order ODE
\begin{equation}\label{scalar_4}j=1.\end{equation}
This constant jerk condition is consistent with recent redshift analysis \cite{planck_2013}.

Let us perform an analogous procedure for the two-component Universe filled with non-relativistic matter with density $M_m/a^3$ and radiation with density $M_r/a^4$ which do not interact with each other. We represent the first Friedmann equation in the form
\begin{equation}\label{friedmann_1}\frac{\dot a^2}{a^2}+\frac k{a^2}=\frac{M_m}{a^3}+\frac{M_r}{a^4},\quad \frac{8\pi G}3=1.\end{equation}
We then differentiate the latter expression twice w.r.t. time to find
\begin{align}
\nonumber \ddot a = & -\frac12\frac{M_m}{a^2}-\frac{M_r}{a^3},\\
\nonumber \dddot a = & \frac{M_m}{a^3}\dot a+3\frac{M_r}{a^4}\dot a.
\end{align}
Using the definitions of the cosmographic parameters, one obtains
\begin{equation}\label{scalar_3_0} q = \frac12 A+B,\quad j = A+3B,\quad A\equiv\frac{M_m}{a^3H^2},\quad B\equiv\frac{M_r}{a^4H^2},\end{equation}
and then
\begin{equation}\label{scalar_4_0} A=-2j+6q,\quad B=j-2q.\end{equation}
In terms of the variables $A$ and $B$ the Friedmann equation (\ref{friedmann_1}) takes on the form
\[\frac k{a^2}=(A+B-1)H^2,\]
or in terms of the cosmographic parameters
\[k=a^2H^2(4q-j-1).\]
Let us check the latter expression for the curvature $k$ in two limiting cases:
\begin{description}
 \item[a)] a flat Universe filled only with non-relativistic matter;
 \item[b)] a flat Universe filled only with radiation.
\end{description}
In the first case (a) one has $B=0$, $q=1/2$, therefore $A=1$ and $k=0$. Note that $j=1$ in this case.

In the second case (b) one has $A=0$, $q=1$, therefore $B=1$ and $k=0$. In this case $j=3$.

The same result can be obtained using the relation (\ref{scalar_3_0}) and it does not require to know value of the deceleration parameter. From the relation (\ref{scalar_3_0}) it follows that in the one-component flat case $A=1$ (a) or $B=1$ (b) respectively and therefore $j=1$ in the first (a) case and $j=3$ in the second (b) case.

Let us now obtain analogue of the equation (\ref{scalar_2}) for the Universe filled with non-relativistic matter (dust) and radiation. In this case
\[s=-3A\left(1+\frac13 q\right)-3B(1+4q),\]
or using the equation (\ref{scalar_4_0}), one finally gets
\[s-6(q-j)+jq=0.\]
It is easy to see that the latter formula reproduces results of the limiting cases. In absence of the radiation $q=1/2$, $j=1$, and therefore $s=-7/2$. This result can be easily verified by direct calculation of the parameter $s$, which can be easily done because $a\propto t^{2/3}$ in this case. In absence of the matter one has instead $a\propto t^{1/2}$, so $q=1$, $j=3$ and $s=-15$.
\section{Averaging deceleration parameter}
Since the DP $q$ is a slowly varying quantity (e.g. $q = 1/2$ for matter-dominated case and $q = -1$ in the Universe dominated by dark energy in form of cosmological constant), then the useful information is contained in its time average value, which is very interesting to obtain without integration of the equations of motions for the scale factor. Let us see how it is possible \cite{lima}. For that purpose let us define average value $\bar q$ of this parameter on time interval $[0,t_0]$ with the expression
\[\bar q(t_0)=\frac1{t_0}\int\limits_0^{t_0}q(t)dt.\]
Making use of the definition of the DP
\[q(t)=-\frac{\ddot a a}{\dot a^2}=\frac{d}{dt}\left(\frac1H\right)-1,\]
it is easy to obtain
\[\bar q(t_0)=-1+\frac1{t_0H_0},\]
or
\begin{equation}\label{background_5_3}t_0=\frac{H_0^{-1}}{1+\bar q}.\end{equation}
As expected, current age of the Universe is proportional to $H_0^{-1}$, but the proportionality coefficient is solely determined by the average value of the DP. It is worth noting that this purely kinematic result depends on curvature of the Universe, nor on number of components filling it, nor on the type of gravity theory used.

The result (\ref{background_5_3}) can be presented in the form
\begin{equation}\label{background_5_4}T=\frac{H^{-1}}{1+\bar q},\end{equation}
where $T$ is age of the Universe. As $\bar q$ is of order of unity, it immediately follows from (\ref{background_5_4}) that the Hubble time $H_0^{-1}$ is s characteristic time scale on any stage of Universe's evolution. Of course the result (\ref{background_5_4}) coincides with that of (\ref{background_2_22}) for the case of constant DP.

Let us dwell on the result (\ref{background_5_3}). For the FLRW metric the age of the Universe in terms of the redshift $z$ is given by the expression
\[t_0=H_0^{-1}\int\limits_0^\infty\frac{dz}{(1+z)\sqrt{\Omega_{m0}(1+z)^3+\Omega_{\Lambda0}}}.\]
Taking SCM values for the parameters $(\Omega_{m0},\Omega_{\Lambda0})$, it is easy to show that
\[\int\limits_0^\infty\frac{dz}{(1+z)\sqrt{\Omega_{m0}(1+z)^3+\Omega_{\Lambda0}}}\approx1.\]
thereby showing that the age parameter for the $\Lambda$CDM model is $H_0t_0\approx1$. According to (\ref{background_5_3}), it is equivalent to $\bar q\approx0$. This result can be also obtained for one-component liquid with $w=-1/3$ (the $K$-matter \cite{lima}), if one recalls the result  (\ref{background_2_15}) for one-component spatially flat Universe. In such a model $a(t)=a_0H_0t$ and $\dot a =const$, therefore it is called "the K-matter coasting model". Another example of coasting model is provided by the Milne empty Universe. However, the coasting empty relativistic solution can be achieved only in the hyperbolic geometry of the Universe, which clearly contradicts the present observations.

\begin{figure}
\includegraphics[width=\textwidth]{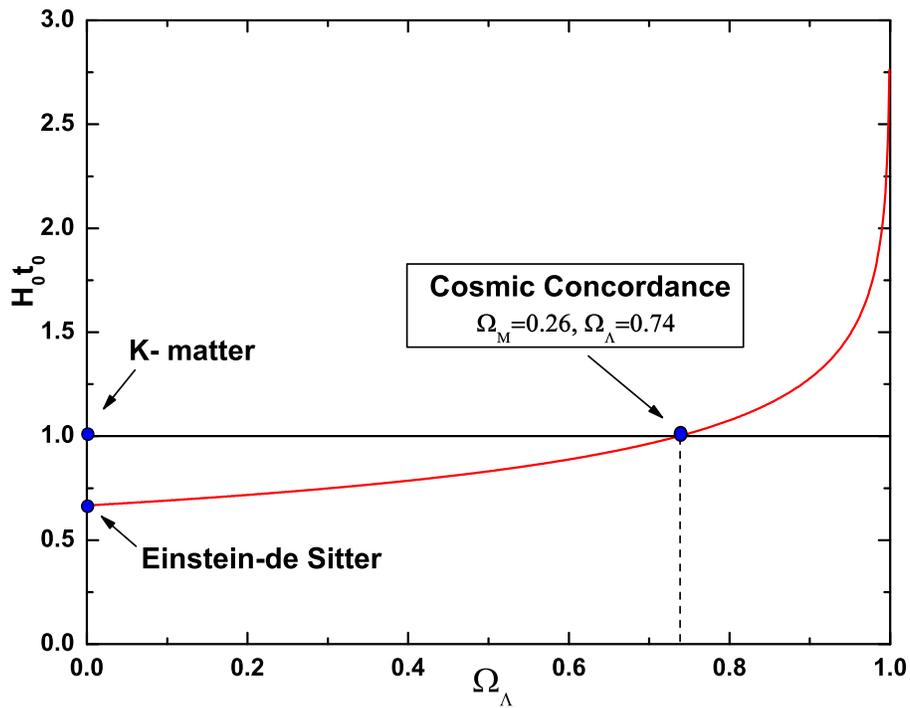}
\caption{\label{f1}Age of the Universe in $\Lambda$CDM and other cosmologies. The present total age of the Universe for the cosmic concordance model ($\Lambda$CDM) is exactly the same as predicted by the flat coasting $K$-matter model, $t_0=H_0^{-1}$ \cite{lima}.}
\end{figure}

\begin{figure}
\includegraphics[width=\textwidth]{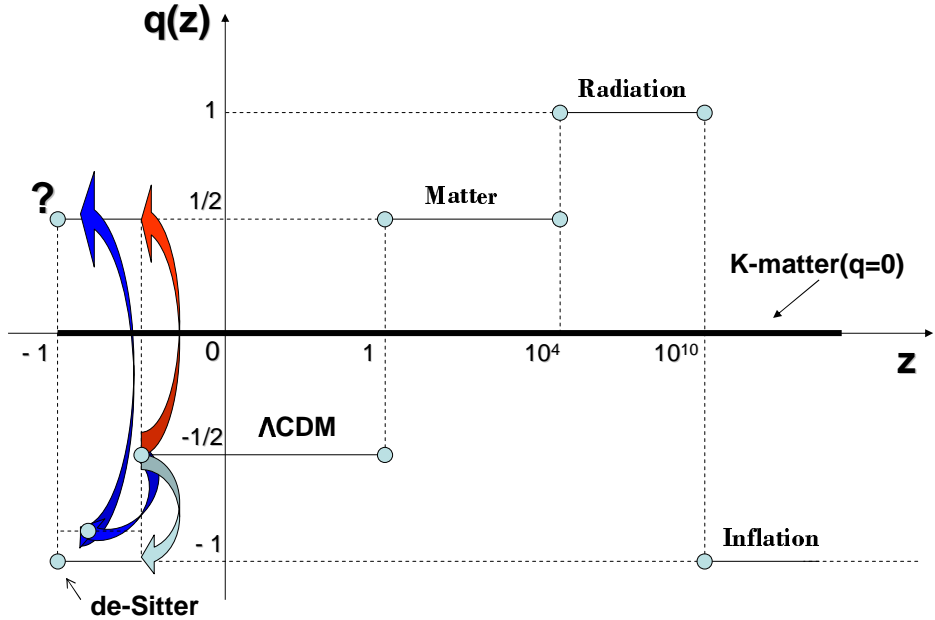}
\caption{\label{f2}The DP as a function of the redshift (the values of $z$ are not in scale).The Universe emerges from an inflationary stage at the infinite past ($z=0$) and evolves to the infinite future ($z=-1$). Abrupt transitions has been assumed for any two subsequent stages. In the actual Universe they are smooth, but, qualitatively, the result is the same, namely, the DP of the Universe seems to be oscillating around the $K$-matter solution ($q=0$). The fate of the Universe is heavily dependent of the next transition in the future. The Universe may decelerate (not necessarily mimicking the dust behavior suggested in the figure) \cite{lima}.}
\end{figure}

In Fig.\ref{f1} \cite{lima} we compare the values of the age parameter $H_0t_0$ for a large set of cosmologies, including $\Lambda$CDM, Einstein - de Sitter, and $K$-matter models. For the SCM  ($\Lambda$CDM), the age of the Universe nowadays is exactly the same one predicted by the coasting $K$-matter model.
Naturally, such a fact may be just an unexpected coincidence. However, the History of Science has already shown that coincidences, mainly in the field of cosmology, deserves a special attention. In terms of the average DP the result $t_0=H_0^{-1}$ means that the average DP $\bar q$ must be identically zero when it is averaged for a long time interval. It should be stressed that it implies equality to zero for the average DP which can be achieved through a cascade of accelerating/deceleration regimes. An example of such a multiple regime cascade is shown on Fig.\ref{f2}.
Let us give some remarks on Fig.\ref{f2}, which presents a hypothetic history of the Universe as a function of the redshift. It starts with the inflation period generated by the cosmological constant ($q=-1$). In order to have the correct age of Universe ($H_0t_0=1$, $\bar q=0$), the average value of the DP after transition to the accelerated expansion regime until the present day must be $-1/2$. Concerning the future (at $z<0$), the dilemma is whether the Universe will transit to an accelerating de-Sitter regime ($q=-1$) as required by the LCDM model, or to a decelerating stage which is predicted by some scalar field and brane-world scenarios (this question will be discussed in more details in the section ''Transitive acceleration''). Thus, if $\bar q=0$ remains true in the near future, the Universe must evolve to a decelerating regime ($q>0$). Naturally, this does not mean that such a transition should be the last one. We cannot exclude possibility of a sequence of transitions with sign change of the DP.

\section{Energy conditions in terms of the deceleration parameter}
Dynamic model-independent constraints on the kinematics of the Universe can further be obtained from the so-called energy conditions \cite{hawking_ellis,carrol,visser3,santos,zhang2}. These conditions, based on quite general physical principles, impose restrictions on the components of the energy-momentum tensor $T_{\mu\nu}$. In choosing a model for the medium (a model, but not the equation of state!), these conditions can be transformed into inequalities restricting the possible values of pressure and density of the medium. In the Friedmann model, the medium is an ideal liquid, for which
\[T_{\mu\nu}=(\rho+p)u_\mu u_\nu - pg_{\mu\nu}\]
where $u_\mu$ is the 4-velocity of the ideal liquid of energy density $\rho$ and pressure $p$,which can be expressed via the scale factor and its derivatives,
\begin{equation}\label{background_6_2}\rho=3M_{Pl}^2\left(\frac{\dot a^2}{a^2}+\frac{k}{a^2}\right),\quad p=-M_{Pl}^2\left(2\frac{\ddot a}{a}+\frac{\dot a^2}{a^2}+\frac{k}{a^2}\right).\end{equation}

In terms of density and pressure the energy conditions take on the form
\begin{align}
\nonumber NEC&\Rightarrow & \rho+p\ge&0, & &\\
\nonumber WEC&\Rightarrow & \rho\ge&0,&\rho+p\ge&0,\\
\nonumber SEC&\Rightarrow & \rho+3p\ge&0,&\rho+p\ge&0,\\
\nonumber DEC&\Rightarrow & \rho\ge&0,&-\rho\le p\le&\rho.
\end{align}
Here, NEC, WEC, SEC, and DEC correspond to the zero, weak, strong, and dominant energy conditions. Because these conditions do not require any definite equation of state for the substance filling the Universe, they impose very simple and model-independent constraints on the behavior of the energy density and pressure. Hence, the energy conditions provide one of the possibilities for explaining the evolution of the Universe on the basis of quite general principles. With expression (\ref{background_6_2}), the energy conditions can be expressed in terms of the scale factor and its derivatives:
\begin{align}
\label{background_6_4} NEC&\Rightarrow & -\frac{\ddot a}{a}+\frac{\dot a^2}{a^2}+\frac{k}{a^2}\ge&0,\\
\nonumber WEC&\Rightarrow & \frac{\dot a^2}{a^2}+\frac{k}{a^2}\ge&0,\\
\nonumber SEC&\Rightarrow & \frac{\ddot a}{a}\le&0,\\
\nonumber DEC&\Rightarrow & \frac{\ddot a}{a}+2\left[\frac{\dot a^2}{a^2}+\frac{k}{a^2}\right]\ge&0.
\end{align}
In the case of a flat Universe, conditions (\ref{background_6_4}) can be transformed into restrictions on the DP $q$:
\begin{align}
\label{background_6_5} NEC&\Rightarrow  q\ge1,\\
\nonumber SEC&\Rightarrow  q\ge0,\\
\nonumber DEC&\Rightarrow  q\le2.
\end{align}
There is no WEC among above conditions because it is always satisfied for arbitrary real $a(t)$.

The conditions (\ref{background_6_5}) considered separately in principle allow a possibility for both decelerated ($q>0$) and accelerated ($q<0$) expansion of the Universe. The constraints by NEC in (\ref{background_6_5}) have clear sense: as follows from the second Friedmann equation, the inequality $\rho+3p\le0$ gives necessary condition for accelerated expansion of the Universe, i.e. the accelerated expansion of the Universe is possible only in presence of components with high negative pressure $p<-\rho/3$. The SEC excludes existence of such components. As a result, $q\ge0$ in this case. At the same time, NEC and DEC are compatible with the condition $p<-\rho/3$ and therefore they allow the regimes with $q<0$.

It worth noting that even before the discovery of the accelerated expansion of the Universe in 1997, Visser \cite{visser4,visser5} already concluded, basing on analysis of the energy conditions, that current observations suggest that SEC was violated sometime between the epoch of galaxy formation and the present. This implies that no possible combination of "normal" matter is capable of fitting the observational data.

Let us make an important remark \cite{gong}. In order to connect the energy conditions with observations, one often needs first to integrate them, and then find the corresponding constraints on some observational variables, such as the distance modulus. Those integral forms can be misleading, and great caution is needed when one interprets them physically.
\chapter{Distance-Deceleration Parameter Relations}
\section{Different cosmological distances in terms of the deceleration parameter}
In cosmology there are many different and equally natural definitions of the notion of ''distance'' between two objects or events, whether directly observable or not. Let us consider \cite{cattoen_visser} several examples of distances between two objects (events) in cosmology.
\paragraph{1. The ''luminosity distance'' $d_L$.}The luminosity distance $d_L$ of an object at redshift $z$ is $d_L=(L/2\pi F)^{1/2}$, where $L$ is the bolometric luminosity for a given object and $F$ is the bolometric energy flux received from that object. The expression for the luminosity distance in a FLRW Universe is
\begin{equation}\label{distance_7_1}
d_L(z)=(1+z)\left\{
\begin{array}{lr}
R\sinh\left[\frac1{H_0R}\int\limits_0^z\frac{dz'}{E(z')}\right], & open\\
H_0^{-1}\int\limits_0^z\frac{dz'}{E(z')}, & flat\\
R\sin\left[\frac1{H_0R}\int\limits_0^z\frac{dz'}{E(z')}\right], & closed
\end{array}
\right.\end{equation}
(Here and below we set $c=1$). Here $R$ is the (comoving) radius of curvature of the open or closed Universe, $E=H/H_0$. The relation (\ref{distance_7_1}) can be rewritten in terms of the DP. For the spatially flat case
\[d_L(z)=(1+z)\int\limits_0^z\frac{dz'}{H(z')}=(1+z)H_0\int\limits_0^z du\exp\left\{-\int\limits_0^u [1+q(v)]d[\ln(1+v)]\right\}.\]

It is useful to give an expression for the luminosity distance up to terms of order of $z^2$
\begin{equation}\label{distance_7_3}d_L=\frac{z}{H_0}\left[1+\left(\frac{1-q_0}{2}\right)z+O(z^2)\right],\end{equation}
where in the spatially flat case
\[q_0=\frac12\sum\limits_i\Omega_{i0}(1+3w_i).\]
It follows from (\ref{distance_7_3}) that for small $z$ the luminosity distance is linearly proportional to the redshift, while the proportionality coefficient equals to inverse value of the Hubble constant. For more distant cosmological objects the luminosity distance in the next order depends on current value of the DP $q_0$, or equivalently on number and type of the components filling the Universe. Expression for the luminosity distance in the next order in the red shift can be presented in the form
\begin{align}
\nonumber d_L(z)= & \frac{cz}{H_0}\left[
1+\frac12(1-q_0)z-\frac16(1-q_0-3q_0^2+j_0)z^2\right.\\
 +&\left.\frac1{24}(2-2q_0-15q_0^2-15q_0^3+5j_0+10j_0q_0+s_0)z^3+O(z^4)\right].
\end{align}
As was expected, the latter decomposition contains the cosmographic parameters defined in terms of the higher order time derivatives of the scale factor ($j\propto d^3a/dt^3$).

As an example we calculate the luminosity distance in the Universe filled with non-relativistic matter. Let us start with the Einstein-de Sitter model ($k=0$). We represent the spatially flat case in the following form (\ref{distance_7_1})
\begin{equation}\label{distance_7_5}d_L=a_0r_1(1+z),\quad r_1=\int\limits_{t_1}^{t_0}\frac{dt}{a(t)}.\end{equation}
In the considered case
\[r_1=\frac{3t_0}{a_0}\left[1-\left(\frac{t_1}{t_0}\right)^{1/3}\right],\quad 1+z=\frac{a_0}{a_1}=\left(\frac{t_0}{t_1}\right)^{2/3},\]
thus
\[r_1=\frac{3t_0}{a_0}\left[1-\left(1+z\right)^{-1/2}\right]=\frac{2}{a_0H_0} \left[1-\left(1+z\right)^{-1/2}\right].\]
We took into account here that in the Einstein-de Sitter model $h_0t_0=2/3$. For the luminosity distance one ultimately obtains
\[d_L=a_0r_1(1+z)=\frac2{H_0}\left(1+z-\sqrt{1+z}\right).\]
\paragraph{2. The ''photon flux distance'' $d_F$.} Introduction of this kind of distance is dictated by the fact, that it is often technologically easier to count the photon flux (photons/sec) than it is to bolometrically measure total energy flux (power) deposited in the detector. Such transition leads to additional factor $(1+z)^{-1/2}$, thus
\[d_F(z)=\frac{d_L}{(1+z)^{1/2}}.\]
\paragraph{3. The ''photon count distance'' $d_P$} is related to the total number of photons absorbed without regard to the rate at which they arrive. Thus the photon count distance contains one extra factor of $(1+z)^{-1}$ as compared to the (power-based) luminosity distance
\[d_P(z)=\frac{d_L}{1+z}.\]
\paragraph{4. The ''angular  diameter  distance'' $d_A$} at last is known to be related to the luminosity distance in the following way
\[d_A(z)=\frac{d_L}{(1+z)^2}.\]
Let us cite expressions for the above considered types of cosmological distance up to terms of order $z^6$ \cite{aviles}:
\paragraph{The luminosity distance}
\begin{align}
    d_L & = \frac{1}{H_0} \cdot \Bigl[ z + z^2 \cdot \Bigl(\frac{1}{2} - \frac{q_0}{2} \Bigr) +
    z^3 \cdot \Bigl(-\frac{1}{6} -\frac{j_0}{6} + \frac{q_0}{6} + \frac{q_0^2}{2} \Bigr) + \nonumber\\
    &+\, z^4 \cdot \Bigl( \frac{1}{12} + \frac{5 j_0}{24} - \frac{q_0}{12} + \frac{5 j_0 q_0}{12} -
    \frac{5 q_0^2}{8} - \frac{5 q_0^3}{8} + \frac{s_0}{24} \Bigr) + \nonumber\\
    &+\, z^5 \cdot \Bigl( -\frac{1}{20} - \frac{9 j_0}{40} + \frac{j_0^2}{12} - \frac{l_0}{120} +
    \frac{q_0}{20} - \frac{11 j_0 q_0}{12} + \frac{27 q_0^2}{40}\nonumber\\ &- \frac{7 j_0 q_0^2}{8} + \frac{11 q_0^3}{8} +
    \frac{7 q_0^4}{8} - \frac{11 s_0}{120} - \frac{q_0 s_0}{8} \Bigr) + \nonumber\\
    &+\, z^6 \cdot \Bigl( \frac{1}{30} + \frac{7 j_0}{30} - \frac{19 j_0^2}{72} + \frac{19 l_0}{720} +
    \frac{m_0}{720} - \frac{q_0}{30} + \frac{13 j_0 q_0}{9}\nonumber\\ &- \frac{7 j_0^2 q_0}{18} + \frac{7 l_0 q_0}{240}
    - \frac{7 q_0^2}{10} + \frac{133 j_0 q_0^2}{48} - \frac{13 q_0^3}{6} + \nonumber\\
    &+\, \frac{7 j_0 q_0^3}{4} - \frac{133 q_0^4}{48} - \frac{21 q_0^5}{16} + \frac{13 s_0}{90}
    - \frac{7 j_0 s_0}{144} + \frac{19 q_0 s_0}{48} + \frac{7 q_0^2 s_0}{24} \Bigr) \Bigr]. \label{distance_7_12}
\end{align}
It should be noted that series expansion of (\ref{distance_7_5}) in terms of $z$ reproduces (\ref{distance_7_12}) for fixed values of the cosmographical parameters, in particular $q=q_0=1/2$.
\paragraph{The photon flux distance}
\begin{align}
    d_F & =  \frac{1}{H_0} \cdot \Bigl[ z - z^2 \cdot \frac{q_0}{2} + z^3 \cdot \Bigl(-\frac{1}{24}
    -\frac{j_0}{6} + \frac{5q_0}{12} + \frac{q_0^2}{2} \Bigr)+ \nonumber\\
    &+\, z^4 \cdot \Bigl( \frac{1}{24} + \frac{7 j_0}{24} - \frac{17 q_0}{48} + \frac{5 j_0 q_0}{12}
    - \frac{7 q_0^2}{8} - \frac{5 q_0^3}{8} + \frac{s_0}{24} \Bigr)  + \nonumber\\
    &+\, z^5 \cdot \Bigl( -\frac{71}{1920} - \frac{47 j_0}{120} + \frac{j_0^2}{12} - \frac{l_0}{120}
    + \frac{149 q_0}{480} - \frac{9 j_0 q_0}{8} + \frac{47 q_0^2}{40} \nonumber\\ &- \frac{7 j_0 q_0^2}{8}
    + \frac{27 q_0^3}{16} + \frac{7 q_0^4}{8} - \frac{9 s_0}{80} - \frac{q_0 s_0}{8} \Bigr)  + \nonumber\\
    &+\, z^6 \cdot \Bigl( \frac{31}{960} + \frac{457 j_0}{960} - \frac{11 j_0^2}{36} + \frac{11 l_0}{360}
    + \frac{m_0}{720} - \frac{1069 q_0}{3840} + \frac{593 j_0 q_0}{288} \nonumber\\ &- \frac{7 j_0^2 q_0}{18}
    + \frac{7 l_0 q_0}{240} - \frac{457 q_0^2}{320} + \frac{77 j_0 q_0^2}{24} - \frac{593 q_0^3}{192} + \nonumber\\ &+ \frac{7 j_0 q_0^3}{4} - \frac{77 q_0^4}{24} - \frac{21 q_0^5}{16}
    + \frac{593 s_0}{2880} - \frac{7 j_0 s_0}{144} + \frac{11 q_0 s_0}{24} + \frac{7 q_0^2 s_0}{24} \Bigr) \Bigr]. \label{distance_7_13}
\end{align}
\paragraph{The photon count distance}
\begin{align}
    d_P & = \frac{1}{H_0} \cdot \Bigl[ z +z^2 \cdot \Bigl( -\frac{1}{2} - \frac{q_0}{2} \Bigr)
    + z^3 \cdot \Bigl( \frac{1}{3} -\frac{j_0}{6} + \frac{2 q_0}{3} + \frac{q_0^2}{2} \Bigr) + \nonumber\\
    &+\, z^4 \cdot \Bigl( -\frac{1}{4} + \frac{3 j_0}{8} - \frac{3 q_0}{4} + \frac{5 j_0 q_0}{12}
    - \frac{9 q_0^2}{8} - \frac{5 q_0^3}{8} + \frac{s_0}{24} \Bigr)  + \nonumber\\
    &+\, z^5 \cdot \Bigl( \frac{1}{5} - \frac{3 j_0}{5} + \frac{j_0^2}{12} - \frac{l_0}{120}
    + \frac{4 q_0}{5} - \frac{4 j_0 q_0}{3} + \frac{9 q_0^2}{5} - \nonumber\\ & -\frac{7 j_0 q_0^2}{8} + 2 q_0^3
    + \frac{7 q_0^4}{8} - \frac{2 s_0}{15} - \frac{q_0 s_0}{8} \Bigr)  + \nonumber\\
    &+\, z^6 \cdot \Bigl( -\frac{1}{6} + \frac{5 j_0}{6} - \frac{25 j_0^2}{72} + \frac{5 l_0}{ 144}
    + \frac{m_0}{720} - \frac{5 q_0}{6} + \frac{25 j_0 q_0}{9} - \nonumber\\ &- \frac{7 j_0^2 q_0}{18}
    + \frac{7 l_0 q_0}{240} - \frac{5 q_0^2}{2} + \frac{175 j_0 q_0^2}{48} - \frac{25 q_0^3}{6} + \nonumber\\
    &+\, \frac{7 j_0 q_0^3}{4} - \frac{175 q_0^4}{48} - \frac{21 q_0^5}{16} + \frac{5 s_0}{18}
    - \frac{7 j_0 s_0}{144} + \frac{25 q_0 s_0}{48} + \frac{7 q_0^2 s_0}{24} \Bigr) \Bigr]. \label{distance_7_14}
\end{align}
\paragraph{The angular diameter distance}
\begin{align}
    d_A & = \frac{1}{H_0} \cdot \Bigl[ z + z^2 \cdot \Bigl( -\frac{3}{2} - \frac{q_0}{2} \Bigr)
    + z^3 \cdot \Bigl( \frac{11}{6} -\frac{j_0}{6} + \frac{7 q_0}{6} + \frac{q_0^2}{2} \Bigr) + \nonumber\\
    &+\, z^4 \cdot \Bigl( -\frac{25}{12} + \frac{13 j_0}{24} - \frac{23 q_0}{12} + \frac{5 j_0 q_0}{12}
    - \frac{13 q_0^2}{8} - \frac{5 q_0^3}{8} + \frac{s_0}{24} \Bigr)  + \nonumber\\
    &+\, z^5 \cdot \Bigl( \frac{137}{60} - \frac{137 j_0}{120} + \frac{j_0^2}{12} - \frac{l_0}{120}
    + \frac{163 q_0}{60} - \frac{7 j_0 q_0}{4} + \frac{137 q_0^2}{40} - \nonumber\\ & -\frac{7 j_0 q_0^2}{8}
    + \frac{21 q_0^3}{8} + \frac{7 q_0^4}{8} - \frac{7 s_0}{40} - \frac{q_0 s_0}{8} \Bigr)  + \nonumber\\
    &+\, z^6 \cdot \Bigl( -\frac{49}{20} + \frac{79 j_0}{40} - \frac{31 j_0^2}{72} + \frac{31 l_0}{720}
    + \frac{m_0}{720} - \frac{71 q_0}{20} + \frac{163 j_0 q_0}{36} - \nonumber\\ & -\frac{7 j_0^2 q_0}{18}
    + \frac{7 l_0 q_0}{240} - \frac{237 q_0^2}{40} + \frac{217 j_0 q_0^2}{48} - \frac{163 q_0^3}{24} + \frac{7 j_0 q_0^3}{4} - \frac{217 q_0^4}{48} - \nonumber\\ & - \frac{21 q_0^5}{16}
    + \frac{163 s_0}{360} - \frac{7 j_0 s_0}{144} + \frac{31 q_0 s_0}{48} + \frac{7 q_0^2 s_0}{24} \Bigr) \Bigr]. \label{distance_7_15}
\end{align}
To avoid problems with the convergence of the series for the highest redshift objects, these relations are recast in terms of the new variable $y = z/(1 + z)$ \cite{cattoen_visser,xu_wang}
\begin{align}
\nonumber H(y) & = H_0\left[\begin{array}{l}
1 + \left( {1 + {q_0}} \right)y + \left( {1 + {q_0} + \frac{1}{2}{j_0} - \frac{1}{2}q_0^2} \right){y^2} \\ \\ +\frac{1}{6}\left( {6 + 6{q_0} + 3{j_0} - 4{q_0}{j_0} - 3q_0^2 + 3q_0^3 - {s_0}} \right){y^3} + \\ \\
 + \left( {1 + {q_0} - 2{q_0}{j_0} + \frac{3}{2}q_0^3 - \frac{1}{2}{s_0}} \right){y^4} + {\rm O}\left( {{y^5}} \right)
\end{array} \right]\\
\nonumber \\
\nonumber {d_L}(y) & = H_0^{ - 1}\left[ \begin{array}{l}
y + \frac{1}{2}\left( {3 - {q_0}} \right){y^2} + \frac{1}{6}\left( {11 - 5{q_0} - {j_0} + 3q_0^2} \right){y^3} + \\ \\
 + \frac{1}{{24}}\left( {50 - 26{q_0} - 7{j_0} + 21q_0^2 + 10{q_0}{j_0} - 15q_0^3 + {s_0}} \right){y^4} + {\rm O}\left( {{y^5}} \right)
\end{array} \right]\\
\nonumber \\
\nonumber {d_A}(y) & = H_0^{ - 1}\left[ \begin{array}{l}
y - \frac{1}{2}\left( {1 + {q_0}} \right){y^2} + \frac{1}{6}\left( {1 - {q_0} + {j_0} - 3q_0^2} \right){y^3} + \\ \\
 + \frac{1}{{24}}\left( { - 2 + 2{q_0} + {j_0} - 3q_0^2 + 10{q_0}{j_0} - 15q_0^3 + {s_0}} \right){y^4} + {\rm O}\left( {{y^5}} \right)
\end{array} \right]
\end{align}
These formulae can become useful if one increases accuracy of the cosmological observations and includes datasets at higher redshifts.

The formula (\ref{distance_7_1}) enables us to find the luminosity distance basing on given function $H(z)$. Let us now solve the inverse problem and find the Hubble parameter as function of the luminosity distance \cite{nesseris_garcia_bellido}. In the spatially flat case differentiation of (\ref{distance_7_1}) results in
\[\frac{d(d_l)}{dz}=\frac{1+z}{H(z)}+\frac{d_L(z)}{1+z}\Rightarrow H(z)=\frac{(1+z)^2}{d'_L(1+z)-d_L},\]
where prime denotes derivative with respect to redshift $z$. Then
\[q(z)=-1\frac{\dot H}{H^2}=-1+(1+z)\frac{H'}{H}=1-\frac{(1+z)^2d''_L(z)}{d'_L(1+z)-d_L},\]
can be rewritten in the form
\[q(N)=-1-\frac{H'(N)}{H(N)}=1+\frac{d''_L(N)+d'_L(N)}{d'_L(N)+d_L(N)}.\]
Here primes denote derivatives with respect to $N\equiv\ln a=-\ln(1+z)$.

For arbitrary geometry
\[q(z)=\frac{1+\omega_K d_L(z)d'_L(z)/(1+z)}{1+\omega_K d_L^2(z)/(1+z)^2}-\frac{(1+z)^2d''_L(z)}{d'_L(z)(1+z)-d_L}.\]
\section{Source counts}
Let us cite without derivation a useful formula for the number of astronomical sources with redshifts in the range $(z,z+dz)$ \cite{shtanov,arbab}
\[dN=\frac{4\pi}{H_0^3}\frac{\left[q_0z+(q_0-1)\left(\sqrt{1+2q_0z}-1\right)\right]^2}{q_0^4(1+z)^6\sqrt{1+2q_0z}}ndz.\]
Here $n$ is the number of sources in a unit proper volume. This equation is applicable to all Friedmann models.
\section{Horizons}
In the present section we analyze connections between the fundamental cosmological parameters with dimension of length: the particle horizon $L_p$, the event horizon $L_e$ and the Hubble's radius $R_H$. The reader is assumed to be acquainted with definition of these quantities. As we shall see, connections between these quantities depend on (and in considerable measure are defined by) the DP, i.e. the character of the cosmological expansion. Moreover, global picture of the observable Universe and its time evolution depend on the DP as well. Indeed, due to the Hubble's law, the galaxies situated on the Hubble's sphere recede with light speed. Velocity of the Hubble's sphere equals to time derivative of the Hubble's radius $R_H=c/H$,
\begin{equation}\label{distance_11_1}\frac{d}{dt}R_H=c\frac{d}{dt}\left(\frac1H\right)= -\frac{c}{H^2}\left(\frac{\ddot a}{a}-\frac{\dot a^2}{a^2}\right)=c(1+q).\end{equation}
As one can see from (\ref{distance_11_1}) the Hubble sphere contracts when $q<-1$, remains stationary when $q=-1$ and expands when $q>-1$. We should stress the Hubble sphere in general does not coincide with the horizons, except when it becomes degenerate with the particle horizon at $q=1$ and with event horizon at $q=-1$.

Let us now analyze \cite{harrison} kinematics of the Hubble's sphere (which defines boundaries of the observable Universe) for different regimes of expansion of the Universe.

In the Universe with decelerated expansion ($q>0$) the Hubble's sphere has velocity exceeding the light speed by the quantity $cq$ and thus it overtakes the galaxies situated on its surface. Therefore the galaxies initially situated outside the Hubble's sphere will initially enter inside. Galaxies at distance $R>R_H$ are later $R<R_H$, and their superluminal recession in the course of time becomes subluminal. The light emitted toward the observer by a galaxy outside the Hubble sphere recedes until it enters inside the sphere. Therefter it starts approaching us and becomes available to observations. Therefore, all decelerating Universes lack event horizons unless they terminate at some future time.

In uniformly expanding Universes $q=0$ the Hubble surface and the galaxies situated on it have equal velocities. Thus number of galaxies in the observable Universe remains constant. Then both particle horizon and event horizon are absent in such Universes.

In case of the accelerating expansion $q<0$ the Hubble sphere has velocity which is less than the light speed by the quantity $q$ and thus it falls behind the galaxies, and therefore number of them decreases inside the Hubble sphere. All accelerating Universes have the property that galaxies at distance $R<R_H$ are later $R>R_H$, and their subluminal recession in the course of time becomes superluminal. Light emitted outside the Hubble sphere recedes from the observer and can never approach the observer. There are events that can never be observed, and such Universe have event horizons.

The above mentioned situation generates the following question of principle. Should we treat as real the galaxies which were initially inside the Hubble sphere and then became unavailable to observation? It is the possibility to check a physical theory experimentally which distinguish the physics from metaphysics. Progress in the experimental technique pushes the boundaries of physics, winning over more and more objects from the metaphysics. Atoms, elementary particles and black holes transformed from metaphysical objects into physical ones during the last century. Being inhabitants of evidently expanding Universe, we face the opposite situation. As well as in decelerated expanding Universe, there are galaxies so distant that no signal from them can be presently detected by terrestrial observer. However if the cosmological expansion accelerates then we recede the galaxies with superluminal velocity. Therefore if their light have not reach us till now then it will never do. Are we treat them as a physical object (because the galaxies are products of the Big Bang as well as ours) or a metaphysical one (due to impossibility to observe them)? Those who believe science fiction to be the realization of unlimited fantasy are quite mistaken. Science fiction is dull and lacks any flight of fantasy compared with cosmology.

Let us now consider several examples which demonstrate kinematics of the Hubble sphere in popular cosmological models. In the power-law models ($a\propto t^n$) we have
\[H=\frac n t, \quad q=\frac{1-n}t, \quad R_H=\frac{ct}n, \quad \dot R_H=\frac c n.\]
In the matter-dominated Universe $n=2/3$, $q=1/2$ the Hubble sphere expands at velocity $3c/2$ and overtakes the comoving galaxies at relative velocity $c/2$. In the radiation-dominated Universe $n=1/2$, $q=1$ the Hubble sphere expands at velocity $2c$ and overtakes the comoving galaxies at relative velocity $c$.

When the scale factor $a(t)\propto\exp(Ht)$, $H=const$ (de Sitter Universe), $q=-1$, the Hubble sphere has a constant radius. Galaxies cross the Hubble surface at velocity $c$. light emitted by these galaxies reaches the observer in the infinite future with infinite redshift. All events outside the Hubble sphere can never be observed, and the Hubble sphere acts as an event horizon.

Using the definitions of the particle horizon and the event horizon one finds
\begin{align}
\nonumber \frac{dL_p}{dt}=&\frac{d}{dt}\left[a(t)\int\limits_0^t\frac{dt'}{a(t')}\right]=L_p(z)H(z)+1;\\
\label{distance_11_3} \frac{dL_e}{dt}=&\frac{d}{dt}\left[a(t)\int\limits_t^\infty\frac{dt'}{a(t')}\right]=L_e(z)H(z)-1.
\end{align}
One can see that observable part of the Universe expands faster than the Universe itself. In other words, the observed fraction of the Universe always increases. Indeed, the particle horizon at distance $L_p$ recedes with velocity $\dot L_p$, while the galaxies at the particle horizon recede at velocity $HL_p$. Hence the horizon overtakes the galaxies with the speed of light $c$.
For Universes of constant $q$
\[\dot R_H=1+q,\quad \dot L_p=1+\frac1q,\quad \frac{L_p}{R_H}=\frac1q.\]
When $q\le0$ no particle horizon exists. The light cone in this case extends to $t=-\infty$. When $0<q<1$ the Hubble sphere lies inside the observe Universe $R_H<L_p$ and bodies receding at velocity $c$ at Hubble sphere have finite redshift. In the radiation-dominated Universe, $q=1$, and the Hubble sphere and observable Universe have the same size.
Differentiating the relations (\ref{distance_11_3}) w.r.t. time, one obtains
\begin{align}
\nonumber \frac{d^2L_p}{dt^2}=&H(1-qHL_p);\\
\label{distance_11_5} \frac{dL_e}{dt}=&-H(1+qHL_e).
\end{align}
Let us cite as an example the expressions for the particle horizon in the single-component Universe filled with non-relativistic matter \cite{shtanov,arbab}. In this case
\[L_p=a_0\int\limits_0^{r_0}\frac{dr}{\sqrt{1-kr^2}} =a_0\int\limits_0^{t_0}\frac{dt}{a(t)}=\frac1{H_0} \left\{\begin{array}{lcr}
2, & k=0, & q_0=1/2;\\
\frac{\arcsin{\frac{\sqrt{2q_0-1}}{q_0}}}{\sqrt{2q_0-1}}, & k=1, & q_0>1/2;\\
\frac{\mathrm{arcsinh}{\frac{\sqrt{1-2q_0}}{q_0}}}{\sqrt{1-2q_0}}, & k=-1, & q_0<1/2.
\end{array}\right.\]
To conclude we would like to stress one more time that interconnection between the Hubble radius and the particle horizon determined by the expansion type (i.e. the DP) plays a principal role in understanding of casual connections between different regions of the Universe. We cite below an extremely bright discussion of this question in \cite{harrison}:

"How, in the Universe of age $t$ can causally connected distances of $L\gg ct$? Let two comoving bodies be separated by a distance $L$ sufficiently small that each lies in the observable Universe of the other. Each body remains thereafter permanently in the other's observable Universe, and the ratio $L/(ct)$ during expansion depends on the behavior of the Hubble sphere.

In a {\it decelerating} Universe the Hubble sphere expands faster than Universe, and a body at distance $L$ either is inside or will soon be inside the Hubble sphere. Hence, any two bodies must eventually recede from each other at subluminal velocity, and the ratio $L/(ct)$ will then decrease in time. In an {\bf\it accelerating} Universe the Hubble sphere expands slower than Universe, and a body at distance $L$ either is outside or will soon be outside the Hubble sphere. Hence, any two bodies must eventually recede from each other at superluminal velocity and the ratio $L/(ct)$ will then increase in time.

How can causally connected distances of $L\gg ct$ exist? The answer is that the Universe passes through a period of accelerated expansion, and causal connections of $L<ct$, established before acceleration, expand superluminally outside the Hubble sphere... A period of accelerated expansion distend all previously established causal connections and increases the distance to the particle horizon."
\chapter{The Effects of a Local Expansion of the Universe}
\section{General Description}
One would expect cosmological expansion to have a significant effect on the dynamics of massive objects \cite{cooperstock_faraoni_vollick,price_romano,carrera_giulin,nandra_lasenby_hobson_1,nandra_lasenby_hobson_2}. Considering the radial motion of a test particle in a spatially-flat expanding Universe it is easy to show that in the Newtonian limit the radial force $F$ per unit mass at a distance $R$ from a point mass $m$ is given by
\begin{equation}\label{local_0_12_1}F=-\frac{m}{R^2}-q(t)H^2(t)R.\end{equation}
Thus, the force consists of the usual $1/R^2$ inwards component due to the central (point) mass $m$ and a cosmological component proportional to $R$ that is directed outwards (inwards) when the expansion of the universe is accelerating (decelerating).The latter formula has evident origin. In order to describe the cosmological expansion one commonly uses two sets of coordinates: the ''physical'' (or Euler) coordinates ($R,\theta,\varphi$) and comoving (or fixed, Lagrangian) coordinates ($r,\theta,\varphi$)\footnote{The angular coordinates are the same for both sets as the cosmological expansion is assumed to be radial}. The two sets are related by the formula $R(t)=a(t)r$. Therefore a point which is fixed w.r.t. cosmological expansion, i.e. with constant coordinates ($r,\theta,\varphi$), has additional radial acceleration
\begin{equation}\left.\frac{d^2R}{dt^2}\right|_{expansion}=R\frac{\ddot a}{a}=-qH^2R.\end{equation}
Let us dwell a bit on the simplest consequences of the additional acceleration appeared in (\ref{local_0_12_1}) due to the expansion of Universe. Consider a Universe which contains no matter (or radiation), but only dark energy in the form of a non-zero cosmological constant $\Lambda$. In this case, the Hubble parameter and, hence, the DP become time-independent and are given by $H=\sqrt{\Lambda/3}$ and $q=-1$. Thus, the force (\ref{local_0_12_1}) also becomes time-independent,
\begin{equation}\label{local_0_12_3}F=-\frac{m}{R^2}+\frac13\Lambda R.\end{equation}
For the case of spatially finite (i.e. non-pointlike) spherically-symmetric massive objects (\ref{local_0_12_3}) is replaced by
\begin{equation}\label{local_0_12_4}F=-\frac{M(R)}{R^2}+\frac13\Lambda R.\end{equation}
where $M(R)$ is the total mass of the object contained within the radius $R$. If the object has the radial density $\rho(R)$ then \[M(R)=\int\limits_0^R4\pi r^2\rho(r)dr.\]
Although the de Sitter background is not an accurate representation of our Universe, the SCM is dominated by dark-energy in a form consistent with a simple cosmological constant.  Even in the simple Newtonian case (\ref{local_0_12_4}), we see immediately that there is an obvious, but profound, difference between the cases $\Lambda=0$ and $\Lambda\ne0$. In the former, the force on a constituent particle of a galaxy or cluster (say) is attractive for all values of $R$ and tends gradually to zero as $R\to\infty$ (for any sensible radial density profile). In the latter case, however, the force on a constituent particle (or equivalently its radial acceleration) vanishes at the finite radius $R_F$ which satisfies \[R_F=[3M(R_F)/\Lambda]^{1/3},\] beyond which the net force becomes repulsive. This suggests that a non-zero $\Lambda$ should set a maximum size, dependent on mass, for galaxies and clusters.

From (\ref{local_0_12_4}), in the Newtonian limit, the speed of a particle in a circular orbit of radius $r$ is given by
\begin{equation}\label{local_0_12_5}V(R)=\sqrt{\frac{M(R)}R-\Lambda R^2}.\end{equation}
from which it is clear that no circular orbit can exist beyond the radius $R_F$.

Let us now consider \cite{nandra_lasenby_hobson_1} a more realistic, time-dependent model for the background cosmological expansion, treating the central massive object as a point mass. In \cite{nandra_lasenby_hobson_2} there was obtained the metric for a point mass $m$ embedded in an expanding cosmological background. In the spatially-flat case, which is a reasonable description of our Universe, and using physical coordinates, the metric takes on the form
\begin{align}\nonumber ds^2=\left[1-\frac{2m}R-R^2H^2(t)\right]dt^2+&2RH(t)\left(1-\frac{2m}R\right)^{-1/2}dRdt-\\ - & \left(1-\frac{2m}R\right)^{-1}dR^2-R^2d\Omega^2.\label{local_0_12_6} \end{align}
This metric is only applicable outside $r>2m$. Nonetheless it is appropriate to use it for our region of interest $r\gg m$. In \cite{nandra_lasenby_hobson_2} there was also obtained the force per unit mass required to keep a test particle at rest relative to the central mass $m$,
\begin{equation}\label{local_0_12_7}F^*=\frac{\frac m{R^2}-RH^2(t)}{\left(1-\frac m R-R^2H^2(t)\right)^{1/2}}+ \frac{RH^2(t)(q(t)+1)\sqrt{\frac{2m}R}}{\left(1-\frac m R-R^2H^2(t)\right)^{3/2}}.\end{equation}
In the region $m\ll r\ll1/H(t)$ taking into account that $F=-F^*$, one finds the radial force on a unit-mass test particle as simply that given in (\ref{local_0_12_1}).

The cosmological force component $-q(t)H^2(t)R$ in the general case is time-dependent. For example, for the standard  LCDM  cosmology, the cosmological force term reverses direction at about $z=z_t\sim0.7$, changing from an inwards directed force at high redshift (decelerating expansion ) to an outwards directed force at low redshift (accelerating expansion). Moreover, the dominance of the dark energy component will increase and so the expansion will tend to the de Sitter background model considered earlier, for which the cosmological force term is time-independent.

The time-dependence of the cosmological force term in the general case leads to the result that the important structure parameter $R_F$ is also time-dependent. For a central point mass, this is given explicitly by
\begin{equation}\label{local_0_12_8}r_F(t)\approx\left[-\frac m{q(t)H^2(t)}\right]^{1/3}.\end{equation}
provided the universal expansion is accelerating, so that $q(t)$ is negative. Time-dependent cosmological force term will act essentially differently (depending on its sign) on the formation and structure of massive objects (galaxies or galaxy clusters) as compared with the simple special case of a time-independent de Sitter background; in the latter case, the expression (\ref{local_0_12_8}) reduces simply to $R_F=(3m/\Lambda)^{1/3}$. At low redshifts, where the dark energy component is dominant, we might expect that the values of $R_F$ obtained using (\ref{local_0_12_8}) will not differ significantly from those obtained assuming a de Sitter background. Clearly, if the expansion is decelerating (forever) then the force due to the central mass $m$ and the cosmological force are both directed inwards and so there is no radius at which the total force vanishes.

The radius $R=R_F$ does not necessarily corresponds to the maximum possible size of the galaxy or cluster: many of the gravitationally-bound particles inside $R_F$ may be in unstable circular orbits. Therefore it is important to know the so-called "outer" radius, which one may interpret as the maximum size of the object, it is the one corresponding to the largest stable circular orbit $R_S$. The radius $R_S$ may be determined as the minimum of the (time-dependent) effective potential for a test particle in orbit about the central mass \cite{hobson_efstathiou_lasenby}.

Remaining within the frames of the Newtonian approximation (a weak gravitational field and low velocities), the equation of motion for the test particle reads
\begin{equation}\label{local_0_12_9} \ddot R\approx-\frac m{R^2}-q(t)H^2(t)R+\frac{L^2}{R^3}.\end{equation}
This is simply the Newtonian radial force expression (\ref{local_0_12_1}) with the inclusion of a centrifugal term. The radius of largest stable circular orbit in this case  is \cite{nandra_lasenby_hobson_1}
\begin{equation}\label{local_0_12_10} R_S(t)=\left[-\frac m{4q(t)H^2(t)}\right]^{1/3}.\end{equation}
Comparing this expression with that for $R_F(t)$ in (\ref{local_0_12_8}), we see that although both radii are  time-dependent, they are related by a constant factor: at any epoch, the radius $R_S(t)$ of the largest stable circular orbit lies a factor $4^{1/3}\approx1.6$ inside the radius $R_F(t)$ at which the total radial force on a test particle is zero and the circular velocity  vanishes. Using $R_S(t)$ instead of $R_F(t)$ enables us to correct our estimates of the maximum possible sizes of galaxies or galaxy clusters \cite{nandra_lasenby_hobson_1}.

If we consider $R_S(t)$ as the maximum possible size of a massive object at cosmic time $t$, and assume that the object is spherically-symmetric and have constant density, then it follows from (\ref{local_0_12_10}) that there exists a time-dependent minimum density (due to maximum size) for objects, given by
\begin{equation}\label{local_0_12_11} \rho_{min}(t)=\frac{3m}{4\pi R^3_S(t)}= -\frac{3q(t)H^2(t)}{\pi}.\end{equation}
As well as (\ref{local_0_12_10}), the latter relation is valid only for $q(t)<0$ (accelerating expansion). For $q(t)>0$, $\rho_{min}(t)=0$. It is easy to see that
\begin{equation}\label{local_0_12_12} \frac{\rho_{min}(t)}{\rho_{crit}(t)}=-8q(t).\end{equation}
For the current moment of time $\rho_{min}(t_0)\approx4.4\rho_{crit}(t_0)$. One more important characteristic -- the minimum fractional density $\delta_{min}\equiv[\rho_{min}(t)-\rho_m(t)]/\rho_m(t0$ contrast -- is closely related to the DP. As $\rho_m(t)=\Omega_m(t)\rho_{crit}(t)$, then
\begin{equation}\label{local_0_12_13} \delta_{min}(t)=-\left[1+\frac{8q(t)}{\Omega_m(t)}\right].\end{equation}
For the current moment of time $(\Omega_{m0}\approx0.3$, $q_0\approx-0.55$) one finds \cite{nandra_lasenby_hobson_1} that $\delta_{min\,0}\approx14$.
\section{"In an expanding Universe, what doesn't expand?"}
The results obtained in the previous section formally related to the objects with cosmological spatial dimensions (such as galaxies and galaxy clusters). Because of importance of those results we would like to follow \cite{price_romano,mil_probleme} in detailed consideration of universal physical aspects of the influence the expansion makes on the local physical objects, taking into account both the non-relativistic and relativistic effects.

It is principal to answer the following question: if space itself is stretching, does this mean that everything in it is stretching? The traditional answer is that the ''bounded'' systems do not take part in the expansion. But if the whole space is stretching, how the bounded systems can avoid even minimal stretching? And what does it mean "to be slightly affected" for a bounded system? Are the bounded system to be stretched with lower rate? Are the weakly bounded systems to be stretched stronger?

Price and Romano \cite{price_romano} made an attempt to answer those questions with the help of a simple model --- a classical ''atom'', composed of a negatively charged ''electron'' of negligible mass, orbiting around a positively charged heavy ''nucleus''. Let us place the classical atom in a homogeneous Universe, where expansion is described by the scale factor $a(t)$. Shall this expansion force the atom to grow in size, i.e. to increase the radius of the first Bohr orbit? Does the answer depend on size of the atom?

For the concrete model under consideration the question whether the atom joins the cosmological expansion can be reformulated in the following way: what coordinate is fixed for the electron -- the physical $R$ or the comoving $r$? Or an intermediate case is realized? If $R=const$, then the cosmological expansion is absent for our model atom. If $r=const$ then the atom completely joins the cosmological expansion.

Price have shown that the answer on the latter question contains both expected and unexpected features. The expected one: it is the relation between the kinematic characteristics of the expansion and the electrostatic interaction that determines which case (the first or the second) is realized. It turned out that sufficiently weakly bounded electron will expand together with the whole Universe, i.e. $r=const$, while more tightly bounded one after initial perturbation of the orbit ignores further expansion and conserves $R=const$. It is unexpected that the intermediate case is absent. As Price notes, the "all-or-nothing" situation takes place.

The model by Price places the massive nucleus at rest in the origin of the spherical coordinate frame ($R,\theta,\varphi$). Position of the electron of mass $m$ orbiting in the equatorial plane $\theta=\pi/2$ is described by the functions $R(t),\varphi(t)$. As there are only radial forces acting on the electron, then its angular momentum $L=mR^2\varphi$ is conserved, and we define the integral of motion for the unit mass electron
\begin{equation}\label{local_12_14} L\equiv R^2\varphi.\end{equation}
In the Newtonnian approximation in presence of the cosmological expansion a radial force per unit mass (teh equation of motion for $R(t)$) takes on the form
\begin{equation}\label{local_12_15} \ddot R=-\frac{C}{R^2}-q(t)H^2(t)R+\frac{L^2}{R^3}.\end{equation}
Here $C\propto qQ$ (the proportionaly coefficient depends on the choice of the unit system, for example in SI units $C=qQ/(4\pi\varepsilon_0 m)$ . Of course the equation (\ref{local_12_15}) up to redefinition of the constant $C$ coincides with the equation (\ref{local_0_12_9}) for a test particle in the gravity field of a galaxy.

The comparative strengths of the electrostatic and cosmological terms in Eq. (\ref{local_12_15}) can be usefully cast as a comparison of time scales for atomic and expansion effects. Let us  define a characteristic atomic time scale $t_{atom}$ as a combination of the parameters ($L$ and $C$) relevant to the electron's motion, $t_{atom}=L^3/C^2$. In the absence of expansion effects the time for the electron to complete a circular orbit is $2\pi t_{atom}$ .

\begin{figure}
\includegraphics[width=\textwidth]{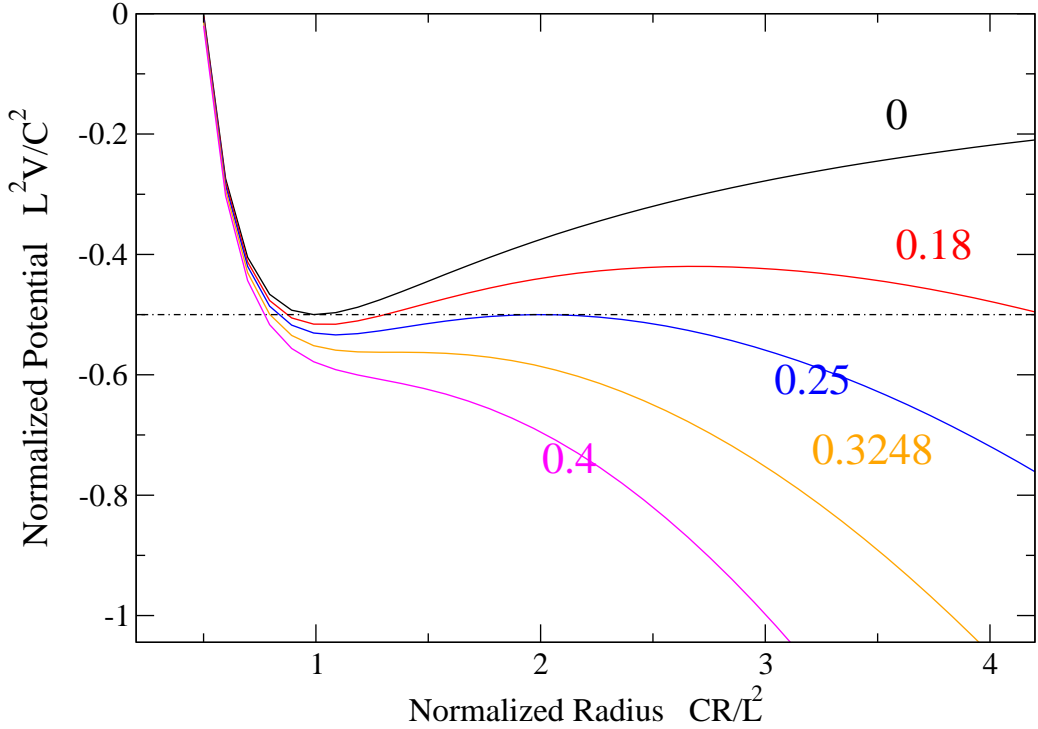}
\caption{\label{local_f_1} Effective potential for exponential expansion. Curves are marked by the value of the parameter $\tau\equiv t_{atom}/t_H$. The curve labeled $0$ is the no-expansion potential, for which $\tau=0$. The dashed line shows the alignment of the minimum for the no-expansion potential with the local maximum of the potential for $\tau=0.25$  \cite{price_romano}.}
\end{figure}

Note that the dynamics of the expansion is coded in functional form of $a(t)$. Choice of this function means hoice of the expansion kinematics, dictated by the dynamics. The concrete choice of $a(t)$ is irrelevant to answer our question of interest, so choosing $a(t)$ one can use the convenience arguments if remaining in frames of realistic expansion laws. So let us consider the simplest type of the accelerated cosmological expansion --- de Sitter cosmology. In this case $a(t)=\exp(t/t_H)$, $q=-1$, $H=const=t^{-1}_H$ and the equation (\ref{local_12_15}) takes on the form
\begin{equation}\label{local_12_16} \ddot R=-\frac{C}{R^2}+\frac{R}{t_H^2}+\frac{L^2}{R^3}.\end{equation}
The first integral of the latter reads
\begin{equation}\label{local_12_17} E\equiv\frac12\dot R^2 + \frac{L^2}{2R^2}-\frac{C}{R}-\frac{R^2}{2t_H^2}.\end{equation}
The electron then feels the effective one-dimensional potential
\begin{equation}\label{local_12_18} V(R)= \frac{L^2}{2R^2}-\frac{C}{R}-\frac{R^2}{2t_H^2}.\end{equation}
Graphical analysis of the potential (\ref{local_12_18}) at different values of the parameters $t_{atom}$ and $t_H$ allows to answer the above posed question about the character of the electron's motion. A graph of the potential for various values $\tau\equiv t_{atom}/t_H$ is shown in Fig.\ref{local_f_1}, where the dimensionless potential $VL^2/C^2$ versus the dimensionless radial distance $CR/L^2$ is shown. Each curve is labeled with the value of the parameter $\tau$ that determines how strongly the cosmological expansion affects the evolution of the atom. The larger the value of $\tau$, the larger is the effect of expansion.

Let us analyze behavior of the electron for different values of the parameter $\tau$. Expansion is absent for the top curve, for which $\tau=0$.  In this case, the electron is always trapped in the potential well, i.e., it is permanently bound. If the electron begins at the bottom of the well ($R=L^2/C$, $E=-C^2/(2L^2)$) it will remain in a circular orbit at that radius for all time. For any larger value of $E$, the electron will orbit in an ellipse. For nonzero values of $\tau$, the potential at large $R$ eventually becomes negative and decreasing, thus representing a dominant outward force. Consequently, an electron at a sufficiently large distance from the nucleus will be driven to an even larger distance. The important question is whether the electron will ever get to this region of dominant outward force. The answer is contained in the shapes of the curves in Fig.\ref{local_f_1}.
We first consider the situation where the electron sits at the bottom of the no-expansion potential well and is "surprised" when the expansion is suddenly turned on. Thus, the electron has energy $E=-C^2/(2L^2)$ and finds itself under the influence of one of the expansion potential curves with $\tau>0$. In this scenario, there is a critical value of $\tau=0.25$, above which the electron will be accelerated outward by the cosmological expansion. As shown by the dashed line in Fig.\ref{local_f_1}, this critical value occurs when the local peak in an expansion potential has the same value as the lowest point in the no-expansion well. For $0<\tau<0.25$, the electron will remain trapped in an approximately elliptical orbit.

A different scenario can also be envisioned. Imagine the electron is sitting at the bottom of an expansion potential well. In this case, the electron will remain at a fixed $R$ (the bottom of the well), assuming such a local minimum actually exists. However, as shown in Fig.\ref{local_f_1}, there is a critical curve that separates potentials that have a local minimum from those that do not. This curve has $\tau=3\sqrt3/16\approx0.324$. This value can be easily obtained using the following arguments.

Extrema of the effective potential in which the test particle moves occur at the $R$-values for which $d^2R/dR^2=0$, namely the solutions of
\begin{equation}\label{local_12_19} -\frac{C}{R^2}+\frac{R}{t_H^2}+\frac{L^2}{R^3}=0.\end{equation}
Consider the function
\begin{equation}\label{local_12_20} y=-t_H^{-2}R^4-CR+L^2.\end{equation}
This polynomial (naturally of $R>0$) has a unique extremum --- a minimum at \[R^*=\left(\frac14Ct_H^2\right)^{1/3}.\]
The existence condition for real roots of the equation (\ref{local_12_19}) reads $y(R^*)\le0$. The critical value of the parameter $\tau$ at which the minimum of the effective potential (\ref{local_12_18}) disappears, corresponds to the condition $y(R^*)=0$ and then $\tau^*=3\sqrt3/16\approx0.324$.

Of course the redefinition $m=C$ and $H=t_H^{-1}$ map all the above obtained results on the case of the test particle in the gravity field of spherically symmetric galaxy considered in the previous section.

Such a qualitative analysis allows us to understand why the atom has an all-or-nothing behavior. The electron either is, or is not, trapped in the potential well. Correspondingly, the atom either expands or does not; there is no "partial expansion" possible. Underlying this graphical understanding is a broader but less precise heuristic explanation of the all-or-nothing effect, an explanation that applies regardless of the specific form of the expansion. The cosmological expansion term \[\frac{\ddot a}{a}R(-qH^2R)\] increases at large physical distances $R$ from the nucleus, whereas the centrifugal and electrical forces both decrease. This implies a sort of instability with respect to expansion. If the electron moves suf?ciently far from the nucleus, the expansion term becomes more important and this pushes the electron even further away.

\begin{figure}
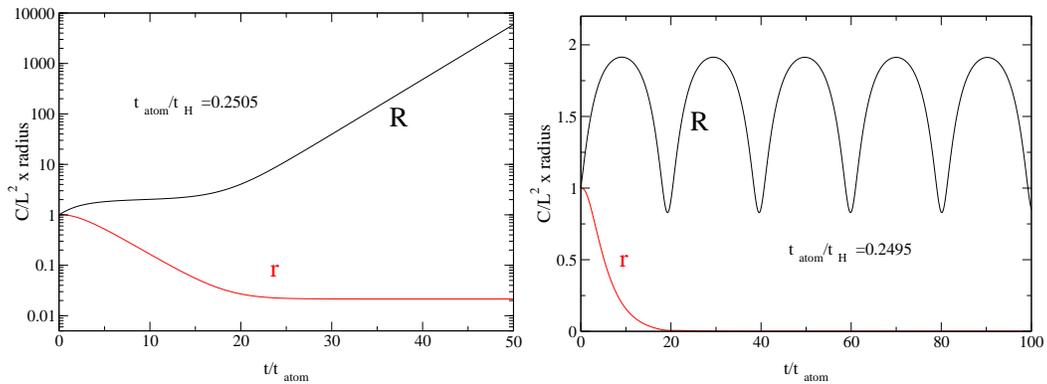

\includegraphics[width=0.5\textwidth]{local_f_2_l.eps}
\includegraphics[width=0.5\textwidth]{local_f_2_r.eps}
\caption{\label{local_f_2} Radial coordinates as a function of time for exponential expansion. On the left is the case for $\tau=0.2505$ for which the electron's comoving radius $r$ remains approximately constant after an initial decrease to about $2\%$ of its initial value. Due to the exponential increase in $a(t)$, the physical radius $R$ grows without bound. On the right is the radial kinetics for a slightly smaller value of $\tau=0.2495$. In this case, the electron remains bound in an approximately elliptical orbit with the physical radius oscillating between values near the original atomic radius. The coordinate radius $R$ in this case falls off exponentially  \cite{price_romano}.}
\end{figure}

We can get yet another viewpoint on the bound/unbound issue by numerically solving Eq. (\ref{local_12_16}). If we start the computation with the electron at the bottom of an expansion well, the results are in agreement with the predictions of the analysis based on Fig.\ref{local_f_1}---the electron remains at fixed $R$. More interesting is the "surprised electron" scenario discussed above (with $E=-C/(2L^2)$). The results, shown in Fig.\ref{local_f_2}, are in accord with the analysis based on Fig.\ref{local_f_1}. For $\tau$ slightly greater than the $0.25$ critical value, the physical radius $R$ of the atom grows exponentially after an initial hesitation. In contrast, for $\tau$ slightly less than this critical value, the electron remains trapped in an approximately elliptical orbit and is not dramatically affected by the exponential expansion.

\section{Tethered Galaxy}
As we have seen above, the unstable character of the cosmological acceleration leads to different observable effects. They can be divided into two groups. The first is due to variation of relative distance between the members of the cosmological expansion regardless the possibility to register this variation. The second is connected with analysis of the additional red shift due to the relative motion. The DP plays a crucial role in both cases. Instead of scholastic disputes on the nature of the red shift which is connected with numerous misunderstandings \cite{davis_lineweaver}, we consider a simple model called the ''tethered galaxy''\cite{davis_lineweaver_webb}.

Let a distant galaxy in the expanding Universe recedes with velocity $V$ given according to the Hubble law by $V=HR$. Imagine that we separated a small test galaxy from the Hubble flow and tied it to the observer galaxy so that the physical (proper) distance between them remains constant. We will not touch the practical realization of this project, as it always can be thought as we apply to the tethered (or now not tied) galaxy a velocity directed toward the observer so that this velocity precisely compensates the hubble velocity of the expansion. After we remove the tie (or switch off the jet engine) there is set the initial condition $\dot R=0$. We define total velocity of the test galaxy as the time derivative of the proper distance
\[V_{tot}=\dot R,\quad \dot R=\dot a r+a\dot r=V_{rec}+V_{pec}.\]

The peculiar velocity $V_{pec}$ is the relative velocity w.r.t. the comoving coordinate frame which the test galaxy is extracted from. This velocity corresponds to our common understanding of velocity and it must be less than the light speed. It is presence of the peculiar velocity which makes the derivative $\dot r\ne0$. The recession velocity $V_{rec}$ of the test galaxy equals to the velocity of the Hubble flow on the distance $R$ and it can be arbitrarily high. We will assume below that the test galaxy has negligibly small mass. According to the construction, the tethered galaxy initially has zero total velocity $\dot R_0$, or
\[V_{rec}=-V_{pec},\quad\Rightarrow\quad \dot a_0r_0=-a_0\dot r_0.\]

With these initial conditions we untie the galaxy and let it sail freely. What will it do then: approach to the observer (in the origin of the coordinate frame), recede or keep the distance?

\begin{figure}
\includegraphics[width=\textwidth]{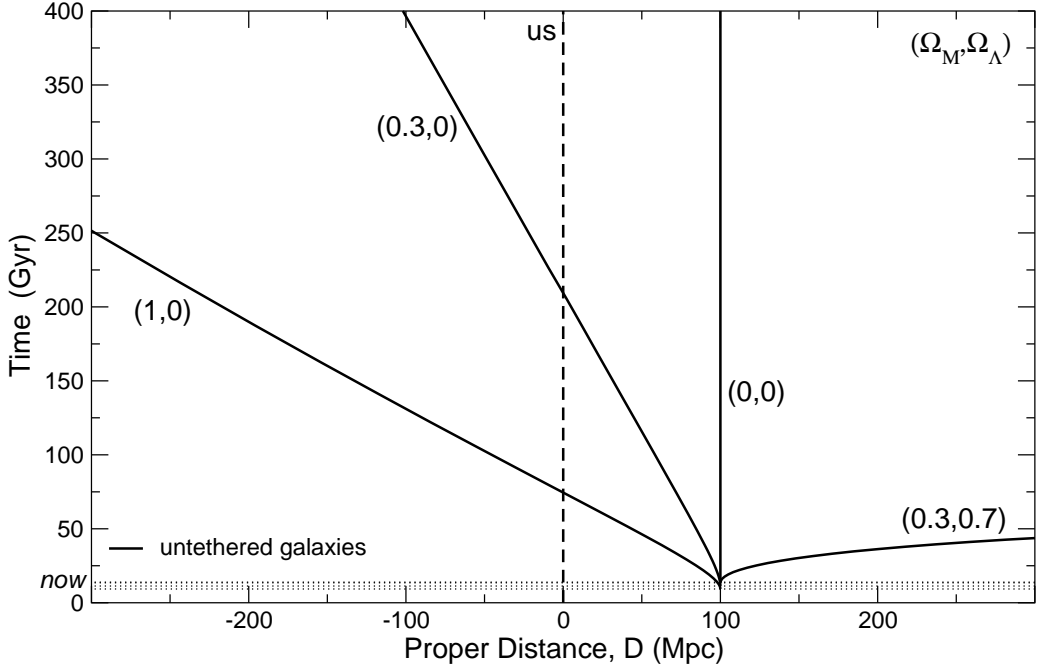}
\caption{\label{local_fig_x1} Graphical interpretation of the tethered galaxy problem.}
\end{figure}

It is principal that the momentum $p$ w.r.t. the comoving reference frame decays as $1/a$. Let us consider the non-relativistic case for simplicity (\cite{davis_lineweaver_webb} contains both non-relativistic and relativistic cases). For the non-relativism $p=mV_{pec}$ and, therefore,
\[V_{pec}=V_{pec0}/a,\quad a\dot r=-\frac{\dot a_0r_0}{a}\Rightarrow r=r_0\left(1-\dot a_0\int\limits_{t_0}^t\frac{dt}{a^2}\right).\]
In order to answer the above posed question one should analyze the time dependence of the proper distance $R$ between the test galaxy and the observer
\begin{equation}\label{local_12_6} R=ar_0\left(1-\dot a_0\int\limits_{t_0}^t\frac{dt}{a^2}\right).\end{equation}
The integral (\ref{local_12_6}) can be evaluated numerically using $dt=da/\dot a$ and $\dot a_0$ obtaijned immediately from the first Friedmann equation with fixed energy content of the Universe. Figure \ref{local_fig_x1} demonstrates solution for four different models: SCM $(\Omega_m,\Omega_\Lambda)=(0,3;0,7)$, empty Universe $(\Omega_m,\Omega_\Lambda)=(0;0)$, Einstein-de Sitter model $(\Omega_m,\Omega_\Lambda)=(1,0)$ and $(\Omega_m,\Omega_\Lambda)=(0,3;0)$. Different behavior in the considered models is determined by the composition of the Universe in each model. It is the composition of the Universe which determines the kinematics of the expansion (accelerated, decelerated or uniform).
For the four cosmological models we untether a galaxy at a distance of $R_0=100Mpc$ with total initial velocity equal to zero and plot its path. In each case the peculiar velocity decays as $1/a$. Its final position depends on the model. In the SCM , accelerating Universe, the untethered galaxy recedes from us as it joins the Hubble flow, while in the decelerating examples $(\Omega_m,\Omega_\Lambda)=(1;0)$ and $(\Omega_m,\Omega_\Lambda)=(0,3;0)$ the untethered galaxy approaches us, passes through our position, and joins the Hubble flow in the opposite side of the sky. In the $(\Omega_m,\Omega_\Lambda)=(0;0)$ model the galaxy experiences no acceleration and stays at a constant proper distance as it joins the Hubble flow.

We have considered above the behavior of the test galaxy in the Universe with fixed DP. The models where the DP can change sign during the time evolutions appear more realistic. Let us make some preliminary notions.

It is not hard to see that the test (untethered) galaxy asymptotically joins the Hubble flow in any cosmological model which is always expanding. Joining of the Hubble flow takes place because
\begin{equation}\label{local_12_7}V_{tot}=V_{rec}+V_{pec}=V_{rec}+\frac{V_{pec,0}}{a}.\end{equation}
At $a\to\infty$ $V_{tot}=\dot R=V_{rec}=HR$, and it is purely Hubble flow. Note that the galaxies join the hubble flow due to growth of the scale factor (or expansion of the Universe). Differentiating (\ref{local_12_7}) w.r.t.time one obtains
\begin{equation}\label{local_12_8}\ddot R = (\ddot a r +\dot a \dot r)-\frac{V_{pec,0}}{a}\frac{\dot a}{a} = (\ddot a r +\dot a \dot r)-V_{pec}\frac{\dot a}{a} = (\ddot a r +\dot a \dot r)-a\dot r\frac{\dot a}{a}=\ddot a.\end{equation}
Using the definition of the DP one finds
\begin{equation}\label{local_12_9}\ddot R=-qH^2R.\end{equation}
An alternative way to obtain (\ref{local_12_9}) is to differentiate the Hubble law. This approach ignores the peculiar velocity and therefore it does not include explicit cancelation of the two terms in (\ref{local_12_9}), which represents more general consideration. The fact that both results are the same, shows that the acceleration of the test galaxy equals that of the comoving galaxy and there is now additional acceleration of the galaxy getting captured by the Hubble flow.

\begin{figure}
\includegraphics[width=\textwidth]{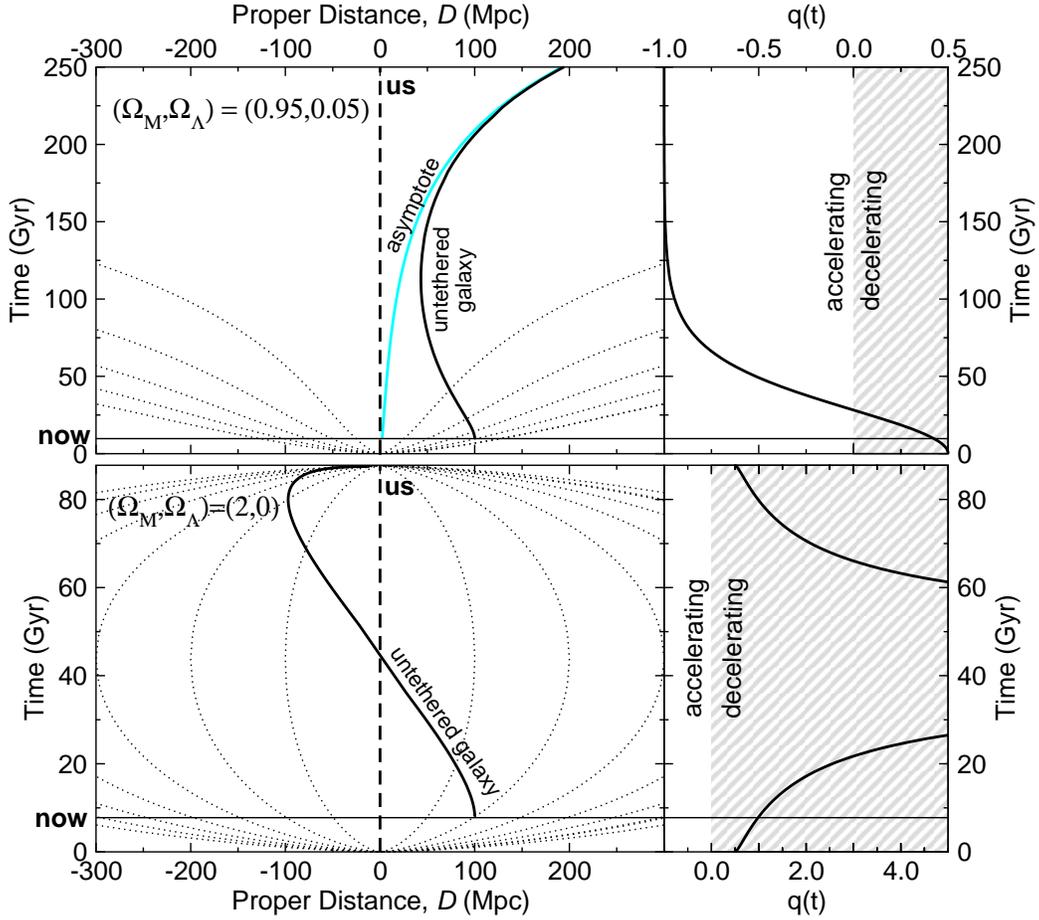}
\caption{\label{local_fig_x2}\cite{davis_lineweaver_webb} Upper panels: $(\Omega_m,\Omega_\Lambda)=(0,95;0,05)$. This particular model shows the effect of $q$ (right panel) on the position of the test galaxy (left panel). Initially $q>0$ and the proper distance to the untethered galaxy decreases.  But $q$ subsequently evolves and becomes negative, reflecting the fact that the cosmological constant begins to dominate the dynamics of the Universe. With $q<0$ the acceleration $\ddot R$ changes sign. This makes the approaching galaxy slow down, stop, and eventually recede. The dotted lines are fixed comoving coordinates. Lower panels: $(\Omega_m,\Omega_\Lambda)=(2;0)$. Universe expands and then recollapses ($\dot a$ changes sign).}
\end{figure}

Recall that $\dot R_0=0$ according to the chosen initial conditions. Therefore it is sign of acceleration $\ddot R$ of the galaxy which determines whether the galaxy approaches us or recedes. The sign in its turn is determined by the sign of the DP according to (\ref{local_12_9}). Equation (\ref{local_12_9}) shows that the sign of $\ddot R$ is determined by the sign of the DP $q$. If the Universe expansion is accelerated ($q<0$) then $\ddot R>0$ and the test galaxy recedes from us. If the Universe expansion is decelerated ($q>0$) then $\ddot R>0$ and the test galaxy approaches us. At last in the case $q=0$ the proper (physical) distance does not change as the test galaxy joins the Hubble flow. The consequence is that in the Universe where the expansion regimes interchange the time dependence of the distance between the test galaxy and the observer is very complicated (Fig.\ref{local_fig_x2}).
\section{Inertia forces in the accelerated expanding Universe}
As we have seen above (\ref{local_12_9}), as the Universe expands, the relative acceleration between two points separated by a distance $R$ is given by $-qH^2R$. If there is a particle with mass $m$ at each of the points, an observer at one of the masses will measure an inertial force on the other mass of
\begin{equation}\label{local_14_1}F_{inert}=-mqH^2R.\end{equation}
Let us estimate magnitude of the force (\ref{local_14_1}) \cite{bamba_capozziello_nojiri_odintsov}. In the Universe filled by non-relativistic matter with zero pressure and dark energy with the state equation $p=w_{de}\rho$ the equation (\ref{local_14_1}) (in current moment of time) can be rewritten as
\begin{equation}\label{local_14_2}F_{inert}=-\frac12mqH^2_0R(1+3w_{de}\Omega_{de}).\end{equation}
For $w_de\sim-1$, $\Omega_{de}\sim0.7$
\begin{equation}\label{local_14_3}F_{inert}=(3\times10^{-36}\sec^{-2})mR.\end{equation}
For a first glance it seems that one can use the obtained estimate for $F_{inert}$ to establish limitations for the state equation parameter for the dark energy \cite{bamba_capozziello_nojiri_odintsov}. The stability condition for a galaxy cluster in the expanding Universe can be rewritten in the form жно представить в виде
\begin{equation}\label{local_14_4}F_{grav}>F_{inert}.\end{equation}
The inertial force (\ref{local_14_1}) can be rewritten as $F_{inert}=-(mR\kappa^2/6)(\rho+3p)$, $\kappa^2=8\pi G$. Accordingly, the Newton gravity which the point particle with mass $m$ suffers is given by
\begin{equation}\label{local_14_5}F_{grav}=G\frac{mM}{R^2}=\frac{4\pi G}{3}mR\rho_{clust}.\end{equation}
It follows then that the stability condition for the galaxy cluster reads
\begin{equation}\label{local_14_6}F_{grav}>F_{inert}.\end{equation}
Since $\rho_{clust}\sim200\rho$, we find, that stability of the cluster is provided by the condition $w_{de}>-70$, which does not pose any noticeable constraint on the state equation parameter for the dark energy.

\end{document}